# A Benchmark of *GW* Methods for Azabenzenes: Is the *GW* Approximation Good Enough?

Noa Marom,[1] Fabio Caruso,[2] Xinguo Ren,[2] Oliver Hofmann,[2] Thomas Körzdörfer,[3] James R. Chelikowsky,[1] Angel Rubio[2,4], Matthias Scheffler,[2] and Patrick Rinke[2]

1. *Center for Computational Materials, Institute of Computational Engineering and Sciences, The University of Texas at Austin, Austin, TX 78712, USA*
2. *Fritz-Haber-Institut der Max-Planck-Gesellschaft, Faradayweg 4-6, 14195, Berlin, Germany*
3. *Computational Chemistry, University of Potsdam, 14476 Potsdam, Germany*
4. *Nano-Bio Spectroscopy Group and ETSF Scientific Development Centre, Universidad del País Vasco, CFM CSIC-UPV/EHU-MPC and DIPC, Avenida Tolosa 72, E-20018 Donostia, Spain*

## Abstract

Many-body perturbation theory in the *GW* approximation is a useful method for describing electronic properties associated with charged excitations. A hierarchy of *GW* methods exists, starting from non-self-consistent $G_0W_0$, through partial self-consistency in the eigenvalues (ev-scGW) and in the Green's function (sc$GW_0$), to fully self-consistent GW (sc*GW*). Here, we assess the performance of these methods for benzene, pyridine, and the diazines. The quasiparticle spectra are compared to photoemission spectroscopy (PES) experiments with respect to all measured particle removal energies and the ordering of the frontier orbitals. We find that the accuracy of the calculated spectra does not match the expectations based on their level of self-consistency. In particular, for certain starting points $G_0W_0$ and sc$GW_0$ provide spectra in better agreement with the PES than sc*GW*.

## Introduction

Many-body perturbation theory in the *GW* approximation[1-5] is a useful method for describing electronic properties associated with charged excitations, such as fundamental gaps,[1, 6] the level alignment at interfaces,[7-18] defect charge transition levels,[19] and electronic transport.[20-27] In this approximation the self-energy, which is the product of the one-particle Green function, *G*, and the screened Coulomb interaction *W*, is taken as the first term in a perturbative expansion in *W*. Owing to the computational cost of fully self-consistent *GW* (sc*GW*) calculations, a range of *GW* schemes, from non-self-consistent to partially self-consistent, have emerged. These constitute a hierarchy of theoretical consistency, in terms of properties that are considered desirable for a generally applicable electronic structure approach, including: (i) independence of the starting point; (ii) satisfaction of conservation laws; and (iii) inclusion of many-body exchange and correlation in the ground-state.

The lowest rung in this hierarchy is the widely used $G_0W_0$ approach, which does not satisfy points (i)-(iii). In this approach, the quasiparticle (QP) excitation energies are obtained from first order perturbation theory as corrections to the eigenvalues from density functional theory (DFT). This amounts to assuming that the orbitals obtained from the underlying DFT calculation mimic the quasi-particle wave-function well enough to treat the difference between the self-energy and the exchange-correlation potential as a small perturbation.[1] Despite the limited validity of the first order perturbative treatment, $G_0W_0$ often yields excellent results. The $G_0W_0$ scheme is the method of choice for the calculation of the QP spectra of solids (see e.g., Refs. 1, 28-40) and has had some notable success in the description of the electronic structure of various organic[10, 13, 41-58] and metal-organic molecules,[57, 59] as well as organic-inorganic interfaces.[8-16] However, the non-self-consistency gives rise to a dependence of the $G_0W_0$ results on the DFT starting point.[28-32, 49, 57-64] Such a dependence may enter both through the DFT orbitals, whose spatial distribution (*e.g.,* the degree of localization/ delocalization) and hybridization may vary, and through the DFT eigenvalues. The starting point dependence of $G_0W_0$ had been demonstrated before for narrow-gap semiconductors, which semi-local functionals predict to be metallic, and for wide-gap semiconductors whose band gaps are severely underestimated by semi-local functionals.[28-32, 61, 63] More recently, the same issue has been addressed for molecular systems.[41, 49, 57-59, 64] It has been suggested that self-interaction errors (SIE), the spurious interaction of an electron with itself,[65] at the DFT level lead to a strong starting point dependence of $G_0W_0$ calculations and to the inadequacy of a semi-local starting point.[30, 57, 59] Indeed, the propagation of SIE from DFT to *GW* has been demonstrated explicitly for one-electron systems.[66-68] In such cases, the inclusion of a fraction of exact exchange (EXX) in hybrid functionals mitigates SIE and often provides a better starting point for $G_0W_0$ calculations.[64]

The second rung in the hierarchy are partially self-consistent *GW* schemes, in which the quasi-particle energies are updated in the construction of the self-energy operator (ev- sc*GW*).[1] The ev-sc*GW* scheme has been shown to yield better results than $G_0W_0$ calculations based on a semi-local starting point for molecules.[41, 49, 69, 70] In the quasi-particle self-consistent *GW* (QP-sc*GW*) method proposed by Faleev, van Schilfgaarde, and Kotani,[71, 72] the one-particle wave-functions are updated by optimizing the starting point with respect to the *GW* perturbation. In this scheme the orbitals are updated by solving the quasiparticle equation with a Hermitian approximation to the *GW* self-energy. This procedure has been applied successfully to a variety of systems, including strongly correlated materials.[71-75] However, both ev-scGW and QP-scGW may still have a considerable starting-point dependence.[40] They also do not satisfy points (ii) and (iii). The third rung in the hierarchy is a partially self-consistent scheme, combining a self-consistent *G* with a non-self-consistent *W* (scGW$_0$).[76] This scheme incorporates many-body effects in the ground state because the Green function is updated (point iii) and satisfies conservation laws (point ii). However, some starting point dependence is still expected, owing to the non-self-consistent W$_0$.

The highest rung in the hierarchy is sc*GW*, in which the Dyson equation is iterated. This is the only method that satisfies properties (i)-(iii). Full self-consistency is the only way to completely eliminate starting point dependence. Another appealing aspect of sc*GW* is that it provides unique total energies and ground state electron densities. Only in the last few years such calculations have been attempted for molecules, owing to their considerable computational cost.[46, 60, 77, 78] Self-consistency has generally yielded improved ionization energies for a set of atoms and molecules, as compared to $G_0W_0$. However, it has been suggested that self-consistency may worsen the description of the QP spectrum,[79, 80] e.g. for the band structure of K and Si.[81] It has also been suggested that sc$GW$ may provide unreliable spectra and total energies for the Hubbard model in the strong correlation regime.[82] Correcting these issues may require going beyond the $GW$ approximation by introducing vertex corrections. Currently, such corrections are in the initial stage of exploration[83-89] and their implementation would come at the price of an even higher computational cost than sc$GW$.

Here, we assess the performance of $GW$ methods, at different levels of self-consistency, for a set of molecules. Benchmark studies of $GW$ methods have typically focused only on the values of the ionization potentials (IP) and/or fundamental gaps of the systems of interest. In contrast, we examine the whole spectrum as well as the predicted character of the frontier orbitals. The symmetry and spatial distribution of the frontier orbitals affect the formation of chemical bonds, photoexcitation, and charge transfer processes. Therefore, in the context of photovoltaics, it is important not only to predict the IP correctly but also to reproduce the character of the highest occupied molecular orbital (HOMO) and the lowest unoccupied molecular orbital (LUMO).

For the purpose of this benchmark study we have chosen to focus on benzene, pyridine, and the diazines: pyridazine, pyrimidine, and pyrazine, illustrated in Figure 1. These molecules are the basic building blocks of polycyclic aromatic hydrocarbons (PAHs), π-conjugated oligomers, and many organic semiconductors and dyes. They embody the basic physics of such systems including the strong correlation effects in aromatic π-systems[82, 87, 90] and the self-interaction effects introduced by the nitrogen lone pairs.[49, 57] Another advantage of these systems is that they are well-characterized experimentally[91-106] and well-studied theoretically by high-level wave-function and Green's function methods.[91, 102, 107-120] We calculate the electronic structure of benzene, pyridine, and the diazines using: (i) semi-local and hybrid DFT (ii) $G_0W_0$, (iii) ev-sc$GW$, (iv) sc$GW_0$, (v) sc$GW$, and (vi) $G_0W_0$ combined with the second-order exchange self-energy (2OX), as an attempt to go beyond the $GW$ approximation. We compare our results to gas phase photoemission spectroscopy (PES) experiments and to reference calculations. We find that the accuracy of the spectral properties of benzene and the azabenzenes does not match the expectations based on the hierarchy established above. In particular, for certain starting points $G_0W_0$ and sc$GW_0$ outperform sc*GW* providing spectra in better agreement with the PES.

## Computational Details

DFT and *GW* calculations were performed using the all-electron numerical atom-centered orbital (NAO) based code, FHI-aims.[58, 121, 122] The NAO basis sets are grouped into a minimal basis, containing only basis functions for the core and valence electrons of the free atom, followed by four hierarchically constructed sets of additional basis functions, denoted by "*tier 1-4*". A detailed description of these basis functions can be found in Ref. 121. Geometry relaxations were performed using the generalized gradient approximation (GGA) of Perdew, Burke, and Ernzerhof (PBE)[123] with a *tier 2* basis set.

A detailed account of the all-electron implementation of *GW* methods in FHI-aims has been given elsewhere.[58, 60] Non-self-consistent $G_0W_0$ and $G_0W_0$+2OX calculations were performed based on the following mean-field starting points: (i) PBE, as a semi-local starting point, (ii) the one-parameter PBE-based hybrid functional (PBEh, also known as PBE0), with 25% of Hartree-Fock (HF) exchange,[124] as a hybrid functional starting point, and (iii) HF. Partially self-consistent ev-sc*GW* and sc$GW_0$ calculations were performed based on PBE and HF starting points. The non-self-consistent and partially self-consistent calculations are denoted as [method]@[starting point], for example, $G_0W_0$@PBE. The $G_0W_0$, $G_0W_0$+2OX, and ev-sc*GW* calculations were conducted with a *tier 4* basis set. For $G_0W_0$, this gives QP energies converged to within 0.1-0.2.[57-59] The sc$GW_0$ and sc*GW* calculations were conducted with a *tier 2* basis set, which has been shown to be adequately converged for scGW.[60]

The orbital self-interaction error (OSIE)[125-127] and orbital shifts[128] were calculated with a local developers version of the PARSEC real-space pseudopotential code,[129, 130] using the PBE functional and Troullier-Martins pseudopotentials.[131]

# Results and Discussion

1. **Density Functional Theory**

Before embarking on computationally intense *GW* calculations it is desirable to predict, based on considerably cheaper semi-local DFT calculations, whether or not strong starting point dependence is expected for non-self-consistent and partially self-consistent schemes. In light of the connection between SIE at the DFT level and the starting point dependence at the $G_0W_0$ level, we begin by assessing the severity of the SIE for benzene, pyridine, and the diazines. For this purpose we use the OSIE, which is evaluated on the basis of the PBE exchange-correlation potential, $v_{xc}^{PBE}$, the Hartree potential, $v_H$, and the orbitals, $\varphi_i$, as follows:

(1) $\quad e_i = \langle \varphi_i | v_H[|\varphi_i|^2] + v_{xc}^{PBE}[|\varphi_i|^2, 0] | \varphi_i \rangle$

Here, $e_i$ is the shift of the Kohn-Sham (KS) eigenvalue $\varepsilon_i^{KS}$ resulting from the SIE in $v_{xc}^{PBE}$. If $e_i$ is similar for all orbitals then the effect of SIE amounts to a shift of the whole KS spectrum by a constant. In such cases, the semi-local spectrum is a good approximation to the ionization energies measured in PES[125-128] as well as a reasonable starting point for $G_0W_0$. In contrast, when $e_i$ of different orbitals varies significantly the semi-local spectrum is distorted by SIE, such that the energy gaps between orbitals and even the ordering of the orbitals are altered.[125-128] In such cases, the semi-local KS eigenvalues and orbitals are not good approximations to the QP energies and wave-function. Figure 2 shows the OSIE relative to that of the HOMO for benzene and the azabenzenes. Visualizations of the HOMO to HOMO-3 orbitals of all molecules are also shown. For all five molecules the OSIE varies widely from one orbital to the next, which does not bode well for semi-local functionals.

The inclusion of a fraction, α, of EXX in hybrid functionals, within a generalized Kohn-Sham (GKS) scheme often mitigates the effect of SIE. This results in one-particle eigenvalues that better approximate QP energies and therefore are typically in better agreement with PES.[57, 59, 125-128] Following Ref. 127, the effect of adding a fraction of EXX may be estimated based on a semi-local DFT calculation. If we neglect the difference

between the GKS and KS orbitals, then the GKS eigenvalues, $\varepsilon_i^{GKS}$, may be approximated by:

$$(2) \quad \varepsilon_i^{GKS} \approx \varepsilon_i^{KS} + \alpha \langle \varphi_i(\mathbf{r}) | v_x^{HF}[n] - v_x^{KS}([n], \mathbf{r}) | \varphi_i(\mathbf{r}) \rangle$$

where the non-local Fock exchange potential, $v_x^{HF}[n]$, is calculated non-self-consistently, using the KS orbitals from the semi-local DFT calculation:

$$(3) \quad v_x^{HF}[n]\varphi_i(\mathbf{r}) = -\sum_{j=1}^{n} \int \frac{\varphi_j(\mathbf{r})\varphi_j^*(\mathbf{r}')}{|\mathbf{r}-\mathbf{r}'|} \varphi_i(\mathbf{r}') d\mathbf{r}'$$

In the following, we use Eq. (2) and (3) to estimate the PBEh and HF eigenvalues for benzene, pyridine, and the diazines, based on a PBE calculation.[132] The estimated eigenvalues are shown in Figure 3. For benzene, the estimated PBEh and HF eigenvalues increase monotonically with the orbital number from the HOMO-10 to the HOMO. Therefore, the addition of any fraction of EXX is not expected to affect the orbital ordering, despite the significant variance in the OSIE. For pyridine and pyrazine, the estimated PBEh eigenvalues increase monotonically but the estimated HF eigenvalues of the HOMO to HOMO-3 deviate from the monotonic trend. In other words, the addition of a large enough fraction of EXX is expected to change the ordering of these orbitals. For pyridazine and pyrimidine, the predicted PBEh eigenvalues of the HOMO to HOMO-3 already deviate from the monotonic trend and the deviation becomes more pronounced for the predicted HF eigenvalues. For these molecules the ordering of the frontier orbitals is expected to change with the addition of a smaller fraction of EXX than for pyridine and pyrazine.

The PBE, PBEh, and HF spectra of benzene, pyridine, pyridazine, pyrimidine, and pyrazine are shown in Figures 4-8, respectively, and compared to gas-phase PES.[94, 98] The DFT spectra are shifted to align the HOMO peak with the corresponding IP, i.e. the total energy difference (ΔSCF) between the neutral and the cation, obtained with the same functional.[133] The calculated spectra are broadened by convolution with a Gaussian (0.4 eV for benzene and 0.3 eV for the azabenzenes) to simulate experimental broadening. We note that the comparison of the calculated spectra to PES is focused mainly on peak positions because cross-section effects in the PES peak intensities are not taken into account here.[134]

For benzene there is no change in the orbital ordering from PBE to PBEh and to HF, as expected from Figure 3. The HOMO and HOMO-1 are degenerate π-orbitals and the HOMO-2 and HOMO-3 are degenerate σ-orbitals. This is in agreement with the existing consensus regarding the character of the frontier orbitals of benzene.[91-93, 97, 103-105, 107, 115-119] The PBE spectrum has the correct spectral shape, but it appears slightly compressed compared to the PES. The PBEh spectrum is in excellent agreement with PES with respect to the spectral shape and the positions of the frontier orbitals. The HF spectrum appears overly stretched with respect to experiment.

Unlike benzene, the frontier orbitals of pyridine and the diazines include *n*-orbitals, i.e., orbitals with contributions from the carbon and nitrogen σ-system as well as from the nitrogen lone-pair.[95] These orbitals are more strongly affected by the SIE and, as shown in Figures 5-8, they tend to drift to lower energies with respect to the π-orbitals as the fraction of EXX is gradually increased. The ordering of the frontier orbitals, obtained with different methods, is summarized in Table 1.

The HOMO and HOMO-1 of pyridine are quite close in energy and the ordering of the *n*- and π-orbitals has been the subject of an ongoing debate in both the experimental and theoretical literature (see also the discussion in Ref. 94 and references therein). Both PBE and PBEh predict the HOMO to be an *n*-orbital and the HOMO-1 and HOMO-2 to be π-orbitals. The *n*-π-π ordering is in agreement with high level wave-function and Green's function based calculations[111-113, 120] and PES experiments.[95, 98-100] The PBE spectrum appears compressed with respect to experiment, yet the spacing between the *n* HOMO and the π HOMO-1 is too large. This may be explained by a shift of the *n*-orbital to higher energies as a result of the SIE associated with the nitrogen lone-pair. The PBEh spectrum is generally in better agreement with the PES peak positions and so is the HOMO—HOMO-1 spacing. It is interesting to note that not all orbitals are affected by the addition of EXX in the same way. The *n*-orbital is shifted to lower energies with respect to the π-orbitals, leading to a reshuffling of the frontier orbitals as more EXX is added. With PBE+35%EXX the *n*-orbital becomes the HOMO-1 and with PBE+80%EXX it becomes the HOMO-2 (not shown for brevity). This orbital ordering is maintained in the HF spectrum. As shown in Figure 5, the addition of an excessive amount of EXX significantly distorts the spectrum: it is overly stretched, the spacing between the frontier orbitals is too large, and the orbital ordering of π-π-*n* is wrong. This picture is consistent with the PBEh and HF eigenvalues estimated based on a PBE calculation.

Figures 6-8 and Table 1 show that the diazines behave similarly to pyridine. For pyridazine (Figure 6), the assignment of the *n*-π-π-*n* character to the HOMO to HOMO-3, respectively, is motivated by PES experiments[95, 98] and Green function based calculations.[113, 114] PBE predicts a wrong orbital ordering of *n*-*n*-π-π and the spectral shape is distorted with the HOMO-2 being very close to the HOMO-1 instead of to the HOMO-3. The addition of 25% EXX in PBEh produces the correct *n*-π-π-*n* orbital ordering and a spectral shape in very good agreement with experiment. The addition of the full amount of EXX in HF causes the *n*-orbitals to drift even lower in energy with respect to the π-orbitals, yielding a wrong ordering of π-π-*n*-*n* and a spectral shape that bears no resemblance to experiment.

For pyrimidine (Figure 7) and pyrazine (Figure 8) the HOMO to HOMO-3 have been assigned to *n*-π-*n*-π orbitals, respectively, based on PES experiments and reference calculations.[96, 98, 99, 101, 102, 106, 109, 110, 113, 114] For both molecules, as for pyridazine, PBE underbinds the *n*-orbitals with respect to the π-orbitals, whereas HF overbinds the *n*-orbitals with respect to the π-orbitals. This leads to an incorrect orbital ordering and a distorted spectral shape. For both pyrimidine and pyrazine, PBEh yields the correct *n*-π-*n*-π ordering and a spectrum in good agreement with experiment.

The changes in the orbital ordering of the diazines with the addition of an increasing amount of EXX are reproduced correctly by the PBE-based estimated PBEh and HF eigenvalues, shown in Figure 3. This demonstrates that the OSIE and the estimated eigenvalues are valuable tools for assessing the effect of SIE for a system of interest, based on a semi-local DFT calculation.

## 2. Non-self-consistent $G_0W_0$

Having demonstrated the effect of the SIE associated with the *n*-orbitals of azabenzenes at the DFT level of theory, we now examine its manifestation for *GW* calculation at different levels of self-consistency, starting with non-self-consistent $G_0W_0$. At this level of approximation, the QP energies are evaluated as perturbative corrections to the KS eigenvalues by solving the linearized QP equation:[1]

(4) $\quad \varepsilon_i^{QP} = \varepsilon_i^{KS} + \langle \varphi_i | \Sigma(\varepsilon_i^{QP}) - v_{xc}^{KS} | \varphi_i \rangle$

where $\Sigma$ is the *GW* self-energy. Within $G_0W_0$, the self-energy is evaluated non-self-consistently, based on KS (or HF) eigenvalues and orbitals. To evaluate the effect of the starting point on the accuracy of the QP energies we use the mean absolute error (MAE), defined as:

(5) $\quad MAE = \sum_{i=1}^{N} |\varepsilon_i^{exp} - \varepsilon_i^{QP}|/N$

with *N* being the number of distinct peaks in the experimental spectra, i.e. the HOMO to HOMO-9 peaks for benzene and the azabenzenes. To quantify the starting point dependence (SPD) we use the mean difference in the $n^{th}$ QP energy obtained from the two extreme starting points in terms of the amount of EXX, i.e. PBE and HF:

(6) $\quad \Delta_{SPD} = \sum_{i=1}^{N} |\varepsilon_{i,HF}^{QP} - \varepsilon_{i,PBE}^{QP}|/N$

The results of these analyses are given in Tables 2 and 3, respectively.

Figures 4-8 show the results of $G_0W_0$ calculations based on PBE, PBEh, and HF starting points for benzene, pyridine, pyridazine, pyrimidine, and pyrazine, respectively. As expected based on the DFT results for benzene, the orbital ordering predicted by $G_0W_0$ is fairly robust to the mean-field starting point, although considerable differences in the QP energies are observed for different starting points. One discrepancy with experiment that particularly stands out in all $G_0W_0$ spectra is that the HOMO-2/HOMO-3 (degenerate for benzene) are too close to the HOMO-4. We also note that the amount of EXX required for obtaining the best agreement with PES for the IP is about 40% (not shown for brevity). However, with this amount of EXX the QP energies of most orbitals, other than the HOMO, are too low compared to the PES. This means that benchmarks

and starting point optimization schemes that focus only on the IP do not necessarily reflect the quality of the whole spectrum.

For the azabenzenes the QP corrections to the GKS eigenvalues, ($\varepsilon_i^{QP} - \varepsilon_i^{GKS}$), are generally more negative for the *n*-orbitals than for the π-orbitals when starting from PBE or PBEh, whereas the trend is inverted for the HF starting point. This leads to a reshuffling of the energy positions of these orbitals in the $G_0W_0$ calculation, as compared to their ordering in the underlying mean-field calculation. For all the azabenzenes, changes in orbital ordering are observed as a function of the fraction of EXX included in the calculation, as reported in Table 1. For pyridine, both the $G_0W_0$@PBE and the $G_0W_0$@PBEh spectra are in agreement with experiment in terms of the spectral shape. In both the *n*-orbital is shifted down in energy with respect to the π-orbitals, as compared to the underlying DFT calculation. Although the spectral shape of the $G_0W_0$@HF spectrum is improved comparing to that of the HF spectrum, a visible distortion is caused by the HOMO-1 and HOMO-2 being nearly degenerate instead of the HOMO and HOMO-1. Only $G_0W_0$@PBE reproduces the reference orbital ordering of *n*-π-π.

For pyridazine and pyrazine, the $G_0W_0$@PBE spectra are qualitatively more similar to the PES in terms of the spectral shape (i.e., the positions of the peaks *relative* to each other) than the $G_0W_0$@PBEh spectra. However, the G$_0$W$_0$@PBEh spectra are still in better quantitative agreement with the PES with respect to the peak positions. For pyrimidine, only the $G_0W_0$@PBEh spectrum is qualitatively similar to the PES. In terms of orbital ordering (see Table 1), for pyridazine, $G_0W_0$ based on all starting points reproduces the reference orbital ordering of *n*-π-π-*n*. For pyrimidine and pyrazine, $G_0W_0$@PBE and $G_0W_0$@PBEh reproduce the reference orbital ordering of *n*-π-*n*-π, whereas $G_0W_0$@HF does not.

Generally, as shown in Table 2, the best agreement with experimental ionization energies is obtained with $G_0W_0$@PBEh, although only $G_0W_0$@PBE reproduces the experimental energy hierarchy for all molecules, as shown in Table 1. Table 3 shows that $G_0W_0$ suffers from a severe starting point sensitivity for all the azabenzenes, with an average difference of approximately 1.38 eV, between HF- and PBE- based $G_0W_0$ ionization energies. The origin of the starting point dependence in $G_0W_0$ can be traced back to differences in the orbitals and orbital energies, used as input for the self-energy calculation. The screening of $W$, being roughly inversely proportional to the occupied-unoccupied transition energies, is severely affected by the (over-) under-estimation of the HOMO-LUMO gap, which generally results in the (under-) over-estimation of screening. For instance, in $G_0W_0$ based on a PBE starting point (smaller gaps) the interaction $W$ is typically "over-screened" whereas, for similar reasons, $W$ is "under-screened" in $G_0W_0$@HF (too large gaps). The (over-) under-screening in turn leads to a systematic error in the description of the excitation spectrum, as exemplified by the overestimation of the QP energies in the $G_0W_0$@HF spectra reported in Figures 4-8. As a result, a $G_0W_0$ calculation based on a DFT starting point with the "right amount" of screening may yield valence spectra in excellent agreement with experiment,[64] as is the case for $G_0W_0$@PBEh. We now proceed to examine to what extent partial self-consistency can alleviate the starting point dependence.

3. **Partial self-consistency in the eigenvalues (ev-sc*GW*)**

It has been suggested that the starting point dependence of the $G_0W_0$ QP energies may be reduced by partial self-consistency in the eigenvalues.[1, 135] In the ev-sc*GW* scheme, the QP equation (eq. 4) is solved iteratively, recalculating the self-energy with QP energies obtained from the previous iteration of the self-consistency loop.[1] The ev-sc*GW* scheme is expected to reduce the overestimation of the screening typically observed in $G_0W_0$ based on semi-local DFT (or the underestimation in the case of HF), as the screened interaction $W$ is now evaluated with occupied-unoccupied transition energies obtained from a *GW* calculation.[41, 49, 69, 70] However, since the orbitals are not updated self-consistently, the starting point dependence cannot be entirely eliminated. As shown in Table 3, self-consistency in the eigenvalues succeeds in significantly reducing the starting point dependence as compared to $G_0W_0$, providing an average difference of 0.4 eV between the QP energies based on HF vs. PBE. The ev-sc$GW$ spectra of benzene, pyridine, pyridazine, pyrimidine, and pyrazine are shown in Figures 9-13, respectively. Generally, ev-sc$GW$@PBE yields improved IPs, as compared to $G_0W_0$@PBE, whereas ev-sc$GW$@HF yields IPs with similar accuracy to $G_0W_0$@HF. We note, however, that evaluating the performance of ev-sc$GW$ based only on the IP and/or HOMO-LUMO gap may give a false impression of an improvement over $G_0W_0$. Examining the entire spectrum reveals that the partial self-consistency in the eigenvalues does not, in fact, lead to a consistent improvement over $G_0W_0$ for benzene and the azabenzenes. As shown in Table 2, the MAE of ev-sc$GW$@HF is similar to that of $G_0W_0$@HF and the MAE of ev-sc$GW$@PBE is worse than that of $G_0W_0$@PBE. For all molecules, the ev-sc$GW$ spectra are overly stretched with respect to the PES, such that large deviations (on the order of 1 eV) from experiment occur for deeper QP states. Moreover, for most systems the orbital ordering deviates from experimental observations (Table 1).

The systematic overestimation of the ev-sc$GW$ QP energies can be understood as a manifestation of the under-screening of the Coulomb interaction $W$, which now resembles $G_0W_0$@HF. Interestingly, the so called, $G_1W_1$ scheme, in which only one eigenvalue update is performed, has been shown to reduce the PBE overscreening and give comparable results to $G_0W_0$ based on a hybrid functional.[136] However, self-consistency ultimately leads to a systematic under-screening in $W$, as manifested by the overall overestimation of the QP energies. Therefore, based on this analysis, partial self-consistency in the eigenvalues cannot be considered as a way to improve the molecular excitation spectrum over $G_0W_0$. The disappointing performance of ev-sc$GW$ emphasizes the importance of updating both eigenvalues and eigenfunctions self-consistently. We therefore proceed to evaluate the performance of the sc$GW_0$ scheme, in which $G$ is computed self-consistently while $W$ remains non-self-consistent.

## 4. Partially self-consistent sc$GW_0$

A partially self-consistent scheme combining a self-consistent G with a non-self-consistent W was first suggested by von Barth and Holm as a way to avoid the computational cost associated with the self-consistency in W and fulfill some of the conservation laws violated by the other schemes discussed above.[76] Within this scheme, G is calculated by iteratively solving the Dyson equation:

(7) $G^{-1} = G_0^{-1} - \Sigma + \Delta v_H$

where G and $G_0$ are the interacting and non-interacting Green functions, respectively, and $\Delta v_\mathrm{H}$ is the change in the Hartree potential. $W_0$ is kept fixed and used to evaluate the self-energy throughout the iterative procedure. The QP energies are then extracted directly from the poles of the self-consistent Green function through the (integrated) spectral function:

(8) $A(\omega) = -i/\pi \, Im\,[Tr\, G(\omega)]$

The spectra of benzene and the azabenzenes, obtained with this sc$GW_0$ scheme, based on PBE and HF starting points, are shown in Figures 9-13. It is clear from a visual inspection of the spectra, as well as from the MAEs in Table 2, that sc$GW_0$@PBE generally yields QP spectra in better agreement with experiment than $G_0W_0$@PBE. In addition, as shown in Table 1, sc$GW_0$@PBE correctly predicts the character of the frontier orbitals of the diazines (though not of pyridine). In contrast to sc$GW_0$@PBE, sc$GW_0$@HF yields overly stretched spectra, similar to ev-sc$GW$@HF. The QP energies are mostly overestimated and considerable deviations from experiment are observed in the whole spectral region for all molecules. The MAE of sc$GW_0$@HF, though somewhat smaller than that of $G_0W_0$@HF and ev-sc$GW$@HF, is considerably larger than that of sc$GW_0$@PBE.

The significant differences between sc$GW_0$@PBE and sc$GW_0$@HF spectra are reflected in the average starting point dependence of 0.70 eV, which is greater than the starting point dependence of ev-sc$GW$. This indicates that the eigenvalues used in the calculation of the screened Coulomb interaction, W, are largely responsible for the starting point dependence of $G_0W_0$. The update of the wave-function (through the iterative calculation of G) reduces the starting point dependence to a lesser extent if the screening is not updated. This means that although the self-consistency in $G$ incorporates many-body (dynamic) correlation effects and exact-exchange in the ground-state, leading to a consistent description of excitations and ground-state, a judicious choice of the DFT starting point is still necessary for $W_0$. Starting from HF leads to underscreening of the Coulomb interaction and to a deterioration of the QP spectra, similarly to $G_0W_0$@HF and ev-sc$GW$@HF. In contrast, sc$GW_0$@PBE can be said to "enjoy the best of both worlds" in the sense that it benefits from an improved treatment of the ground-state electronic structure through the self-consistency in G, while preserving the PBE screening in the non-self-consistent $W_0$. Due to the underestimation of the HOMO-LUMO gap in PBE-based calculations, the resulting screened Coulomb interaction is slightly overscreened. It has been argued that this effect might mimic the missing vertex corrections (i.e. the electron-hole contribution to the dielectric function), which explains the success of sc$GW_0$@PBE.[137, 138] We expect other partially self-consistent approaches in which the one-particle wave-functions are updated through the approximate solution of the QP equation (e.g. the quasi-particle self-consistent $GW$ approach,[62, 72] or $G_0W_0$ based on the Coulomb-hole plus screened exchange (COHSEX) approximation[139]) to yield QP spectra of similar quality to sc$GW_0$@PBE. We now turn to fully self-consistent $GW$ to evaluate the effects of the self-consistent computation of the screening on the spectral properties of benzene and the azabenzenes.

## 5. Fully self-consistent *GW* (sc*GW*)

As we have demonstrated above, the performance of non-self-consistent and partially self-consistent $GW$ schemes is contingent on choosing a good starting point. Therefore, the only way to eliminate the starting point dependence completely and to truly evaluate the quality and validity of the $GW$ approximation itself is full self-consistency. In sc$GW$, the Dyson equation for G (equation 7) is solved self-consistently, fully updating all the diagonal and non-diagonal matrix elements of $G$ and $\Sigma$, without introducing approximations in the computation of the screened Coulomb interaction. Moreover, within the all-electron sc$GW$ implementation in FHI-aims, the core-valence screening is also updated self-consistently, leading to ground and excited state properties independent of the starting point.[60] The QP energies are obtained from the poles of the spectral function (eq. 8). A complete account of the implementation of sc$GW$ in FHI-aims is given in Ref. 60.

The sc$GW$ spectra of benzene and the azabenzenes are shown in Figures 9-13. The sc$GW$ results are insensitive to the starting point and we obtain the same final spectrum regardless of whether the calculation is started from PBE or from HF. Overall, sc$GW$ provides a better description of the QP energies than $G_0W_0$@PBE, $G_0W_0$@HF, ev-sc$GW$, and $GW_0$@HF for the systems considered here. However, its performance is not as good as one might expect, as it fails to reproduce some important qualitative features of the spectra, such the spectral shape and the ordering of the frontier orbitals of pyridine, pyrimidine, and pyrazine (see Table 1). An appropriate choice of the starting point for $G_0W_0$ or sc$GW_0$, may correctly reproduce these features, outperforming sc$GW$. This is reflected by the lower MAE (Table 2) of $G_0W_0$@PBEh and sc$GW_0$@PBE. Interestingly, the sc$GW$ spectra resemble those of the HF-based schemes with respect to the orbital ordering in the frontier region. In this respect, the non-self-consistent $G_0W_0$@PBEh and the partially self-consistent sc$GW_0$@PBE seem to capture or otherwise compensate for the missing correlation in sc$GW$. This is possibly due to a fortunate error cancellation, whereby the overscreening in the DFT based $W_0$ compensates for neglecting the vertex function. Now, one may ask whether including additional Feynman diagrams would lead to an improved description of the correlation energy, resulting in better agreement with the PES. We therefore examine such a way of going beyond the $GW$ approximation.

## 6. $G_0W_0$ with second order exchange ($G_0W_0$+2OX)

In physical terms, the correlation part of the *GW* self-energy corresponds to higher-order direct scattering processes. Higher-order exchange processes, however, are neglected. The simplest correlation method that treats direct and exchange interactions on an equal footing is second-order Møller-Plesset perturbation theory (MP2), where both direct and exchange processes are taken into account up to second-order. It has been suggested that adding the second-order exchange (2OX) diagram to the self-energy may correct the self-screening errors in *GW*, whose effect, like that of SIE, is more significant for localized states.[89] For the direct term, it is essential to sum over the so-called ring diagrams to infinite order to avoid divergence for systems with zero gaps. In contrast, for exchange-type interactions, the second-order exchange diagram, illustrated in Figure 14 is the dominant contribution to the self-energy and neglecting the higher order diagrams does not lead to a divergence. Thus, the *GW*+2OX scheme, suggested here, is a simple practical correction to the *GW* approximation. Within this scheme, the self-energy is written as:

(9) $\quad \Sigma^{GW+2OX} = \Sigma^{GW} + \Sigma^{2OX}$

where $\Sigma^{2OX}$ is given in terms of the Green's function and the bare Coulomb interaction, $v$, as:[140]

(10) $\quad \Sigma^{2OX}(1,2) = i \int d3d4\, G(1,3)G(3,4)G(4,2)v(1,4)v(3,2)$.

Here, the numbers represent combined space-time coordinates, e.g., $1=(r_1,t_1,\sigma_1)$. The one-particle Green's function, $G_0$, is used to evaluate the 2OX self-energy, which reduces equation 8 to an expression involving only single-particle orbitals and eigenvalues:[58, 141]

(11) $\quad \Sigma^{2OX}_{n\sigma}(\omega) = (np,\sigma|la,\sigma)(pl,\sigma|an,\sigma)\left[\dfrac{\theta(\varepsilon_F-\varepsilon_{p\sigma})}{\omega+\varepsilon_{a\sigma}-\varepsilon_{l\sigma}-\varepsilon_{p\sigma}-i\eta} + \dfrac{\theta(\varepsilon_{p\sigma}-\varepsilon_F)}{\omega+\varepsilon_{l\sigma}-\varepsilon_{a\sigma}-\varepsilon_{p\sigma}+i\eta}\right]$

where $\sigma$ is a spin index, $\Theta(x)$ is the Heaviside step function, $\varepsilon_F$ is the Fermi level, $\eta$ is a positive infinitesimal, and *(np,σ|la,τ)* is the two-electron Coulomb repulsion integral over single-particle orbitals:

(12) $$(np,\sigma|la,\tau) = \iint drdr' \frac{\varphi_{n\sigma}^*(r)\varphi_{p\sigma}(r)\varphi_{l\tau}^*(r)\varphi_{a\tau}(r)}{|r-r'|}$$

While the *GW*+2OX scheme is physically motivated and conceptually appealing, its usefulness can only be judged *a posteriori*, based on its performance, which we assess here at the $G_0W_0$ level.

Figures 15-19 show the $G_0W_0$+2OX spectra of benzene and the azabenzenes, based on different starting points, compared to the PES experiments. Because the $G_0W_0$+2OX scheme is non-self-consistent, a significant starting point dependence of 0.8 eV on average is observed (Table 3). This starting point dependence is smaller than that of $G_0W_0$ but larger than that of the partially self-consistent schemes.

Overall, adding the second-order exchange at the $G_0W_0$ level is not worthwhile. It does not alleviate the starting point dependence and yields worse agreement with experiment in terms of the spectral shape (for all molecules) and the ordering of the frontier orbitals of pyridine, pyrimidine, and pyrazine. This is possibly a result of using the bare, rather than the screened, Coulomb interaction in the 2OX self-energy. Second-order screened exchange (SOSEX), in which one of the bare Coulomb line is replaced by a *dressed* line (i.e., *v* is replace by *W*), was proposed as a possible correction of the self-screening error that affects the *GW* self-energy. In particular, the SOSEX self-energy cancels exactly the self-screening in the one-particle case[142] and is therefore expected to improve the spectral properties of molecules and solids, at the price of a considerably higher computational cost. This calls for further investigation of vertex corrections, which will be pursued in the future.

## Conclusion

We have conducted a benchmark study of the performance of *GW* methods, at different levels of self-consistency, for benzene and azabenzenes, as a set of representative organic molecules. The quality of the calculated spectra was evaluated based on a comparison to PES experiments, in terms of all valence peak positions, as well as the frontier orbital character.

First, we demonstrated that it is possible to assess whether a significant starting point dependence is expected for non-self-consistent and partially self-consistent schemes, based on two simple tests at the semi-local DFT level: (i) the orbital self-interaction error (OSIE) as a measure of the severity of the self-interaction error for the system of interest (ii) estimated hybrid eigenvalues show to what extent the addition of EXX changes the orbital ordering and the shape of the spectrum. These tests revealed that for the azabenzenes, which possess nitrogen lone-pair orbitals, the effects of SIE and of the addition of EXX are considerably more dramatic than for benzene with respect to the ordering of the frontier orbitals.

A significant starting point dependence was found for all the non-self-consistent and partially self-consistent *GW* schemes. The best agreement with the PES was obtained with $G_0W_0$@PBEh and sc$GW_0$@PBE. Unlike partial self-consistency in *G*, partial self-consistency in the eigenvalues was found to cause underscreening and deterioration of the spectra, regardless of the starting point. Although in some cases ev-sc*GW* improved the IP with respect to $G_0W_0$, the ev-sc*GW* spectra generally appeared overstretched as compared to experiment.

Due to underscreening, the spectra obtained from Hartree-Fock based calculations are distorted, and systematically overestimate the QP energies for all perturbative and partially self-consistent schemes analyzed in the present work. We therefore conclude that HF is generally inadequate as starting point for the computation of spectral properties of molecules. Interestingly, no type of partial self-consistency improves on *$G_0W_0$*@HF.

Full-self consistency succeeded in eliminating the starting point dependence, providing an unbiased reference for the performance of the *GW* approximation for benzene, pyridine and the diazines. The sc*GW* spectra improve the quasi-particle energies as compared to PBE and HF based $G_0W_0$, all ev-sc*GW* calculations, and sc$GW_0$@HF. However, for the systems studied here, $G_0W_0$@PBEh and sc$GW_0$@PBE outperformed sc*GW*. In this respect, the success of $G_0W_0$@PBEh may be explained by a fortunate error cancellation, whereby the "right amount" of DFT overscreening compensates for neglecting the vertex function. Applying similar considerations, *$GW_0$*@PBE may be said to "enjoy the best of both worlds", as it benefits from an improved treatment of the correlation through the self-consistency in *G*, while preserving the PBE overscreening in the non-self-consistent $W_0$.

As an initial foray into the land beyond *GW*, we examined the effect of adding the second-order exchange contribution to the self-energy at the $G_0W_0$ level. The resulting $G_0W_0$+2OX spectra were in worse agreement with experiment than the corresponding $G_0W_0$ spectra and seemed overstretched to an even greater extent than the ev-sc*GW* spectra. This may be a result of using the bare, rather than the screened, Coulomb interaction in the 2OX self-energy. This and the effect of adding the 2OX self-energy to sc*GW* will be investigated in future work.

## Acknowledgements

Work at UT-Austin was supported from DOE grant DE-SC000887. Computational resources were provided by the National Energy Research Scientific Computing Center (NERSC), the Texas Advanced Computing Center (TACC), and the Argonne Leadership Computing Facility (ALCF). A.R. acknowledges financial support from MEC (FIS2011- 65702-C02-01), Grupos Consolidados UPV/EHU del Gobierno Vasco (IT-319-07), and the European Research Council (ERC-2010-AdG - No. 267374).

Table 1: Summary of the frontier orbital ordering obtained for azabenzenes with different DFT and *GW* methods. Agreement with the reference is indicated in boldface. a) Refs. 94, 95, 98-100, 111-113, 120 b) Refs. 95, 98, 113, 114 c) Refs. 96, 98, 99, 101, 102, 106, 109, 113, 114 d) Refs. 91, 98, 99, 101, 110, 113

|  | pyridine | pyridazine | pyrimidine | Pyrazine |
|---|---|---|---|---|
| **Reference** | ***n*-π-π**[a] | ***n*-π-π-*n***[b] | ***n*-π-*n*-π**[c] | ***n*-π-*n*-π**[d] |
| **PBE** | ***n*-π-π** | *n*-*n*-π-π | *n*-*n*-π-π | ***n*-π-*n*-π** |
| **PBEh** | ***n*-π-π** | ***n*-π-π-*n*** | ***n*-π-*n*-π** | ***n*-π-*n*-π** |
| **HF** | π-π-*n* | π-π-*n*-*n* | π-*n*-π-*n* | π-*n*-π-*n* |
| $G_0W_0$@PBE | ***n*-π-π** | ***n*-π-π-*n*** | ***n*-π-*n*-π** | ***n*-π-*n*-π** |
| $G_0W_0$@PBEh | π-*n*-π | ***n*-π-π-*n*** | ***n*-π-*n*-π** | ***n*-π-*n*-π** |
| $G_0W_0$@HF | π-π-*n* | ***n*-π-π-*n*** | *n*-π-π-*n* | π-*n*-π-*n* |
| ev-sc*GW*@PBE | π-*n*-π | ***n*-π-π-*n*** | *n*-π-π-*n* | ***n*-π-*n*-π** |
| ev-sc*GW*@HF | π-*n*-π | ***n*-π-π-*n*** | *n*-π-π-*n* | π-*n*-π-*n* |
| sc*GW*$_0$@PBE | π-*n*-π | ***n*-π-π-*n*** | ***n*-π-*n*-π** | ***n*-π-*n*-π** |
| sc*GW*$_0$@HF | π-*n*-π | ***n*-π-π-*n*** | *n*-π-π-*n* | π-*n*-π-*n* |
| sc*GW* | π-*n*-π | ***n*-π-π-*n*** | *n*-π-π-*n* | π-*n*-π-*n* |
| $G_0W_0$@PBE+2OX | π-*n*-π | ***n*-π-π-*n*** | *n*-π-π-*n* | π-*n*-π-*n* |
| $G_0W_0$@PBEh+2OX | π-*n*-π | ***n*-π-π-*n*** | *n*-π-π-*n* | π-*n*-π-*n* |
| $G_0W_0$@HF+2OX | π-*n*-π | ***n*-π-π-*n*** | π-*n*-π-*n* | π-*n*-π-*n* |

Table 2: MAE (eq. 5) in eV for the QP energies of benzene and the azabenzenes obtained with different $GW$ methods with respect to PES.[94, 98]

|  | Benzene | Pyridine | Pyridazine | Pyrimidine | Pyrazine | Average |
|---|---|---|---|---|---|---|
| **$G_0W_0$@PBE** | 0.39 | 0.34 | 0.36 | 0.4 | 0.4 | 0.38 |
| **$G_0W_0$@PBEh** | 0.18 | 0.19 | 0.16 | 0.12 | 0.22 | 0.17 |
| **$G_0W_0$@HF** | 1.07 | 1.11 | 1.06 | 0.99 | 1.01 | 1.05 |
| **sc$GW$** | 0.45 | 0.31 | 0.25 | 0.25 | 0.28 | 0.31 |
| **sc$GW_0$@HF** | 0.92 | 0.95 | 0.93 | 0.88 | 0.88 | 0.91 |
| **sc$GW_0$@PBE** | 0.35 | 0.27 | 0.27 | 0.22 | 0.24 | 0.27 |
| **ev-sc$GW$@HF** | 0.99 | 1.00 | 1.12 | 0.91 | 0.91 | 0.99 |
| **ev-sc$GW$@PBE** | 0.53 | 0.57 | 0.58 | 0.66 | 0.52 | 0.57 |

Table 3: The starting-point dependence (eq. 6) in eV obtained at different levels of $GW$ self-consistency for benzene and the azabenzenes.

|  | $G_0W_0$ | ev-sc$GW$ | sc$GW_0$ | sc$GW$ | $G_0W_0$+2OX |
|---|---|---|---|---|---|
| **Benzene** | 1.32 | 0.41 | 0.87 | 0.0 | 0.72 |
| **Pyridine** | 1.37 | 0.41 | 0.64 | 0.0 | 0.75 |
| **Pyridazine** | 1.42 | 0.42 | 0.66 | 0.0 | 0.77 |
| **Pyrimidine** | 1.40 | 0.40 | 0.66 | 0.0 | 0.94 |
| **Pyrazine** | 1.38 | 0.40 | 0.70 | 0.0 | 0.80 |
| **Average** | 1.38 | 0.40 | 0.70 | 0.0 | 0.80 |

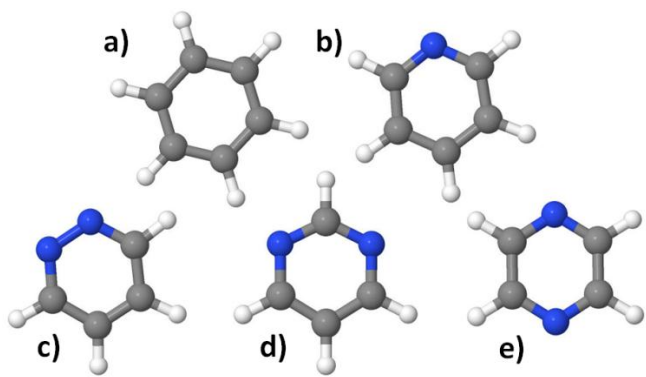

Figure 1: Schematic illustration of the azabenzenes molecules studied here: a) benzene, b) pyridine, c) pyridazine, d) pyrimidine, and e) pyrazine.

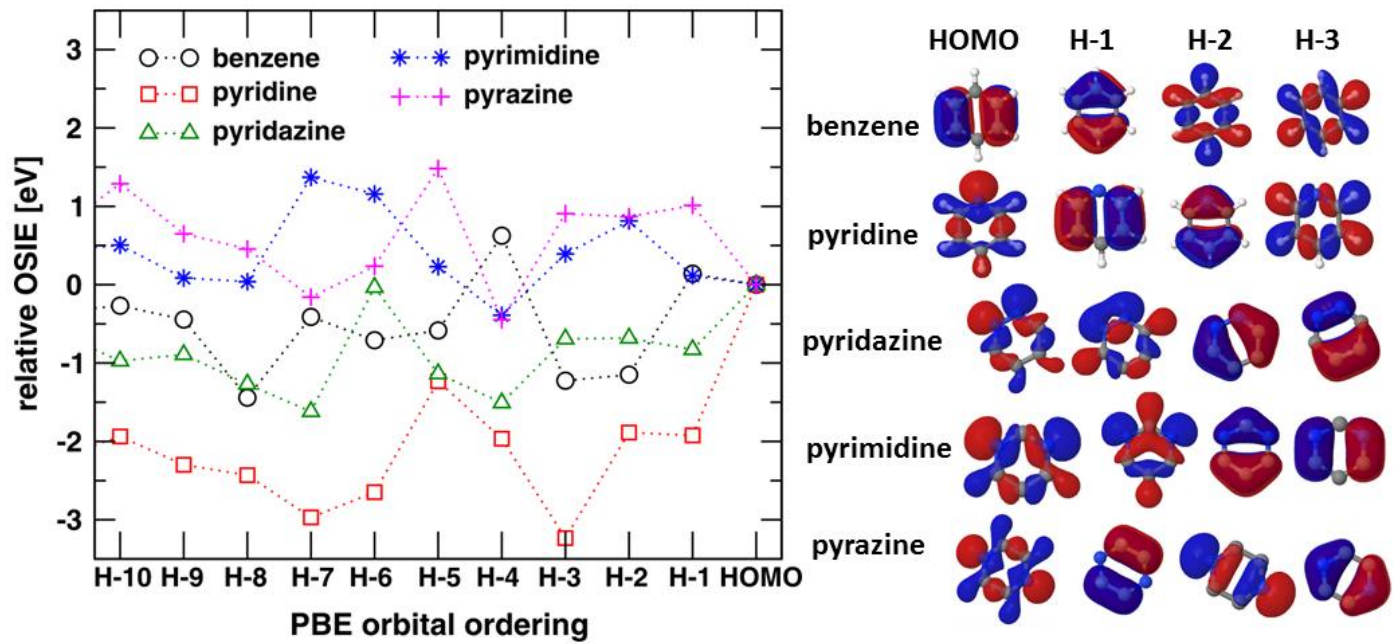

Figure 2: Relative OSIE with respect to the HOMO for benzene, pyridine, pyridazine, pyrimidine, and pyrazine. Visualizations of the HOMO to HOMO-3 orbitals at the PBE ordering are also shown.

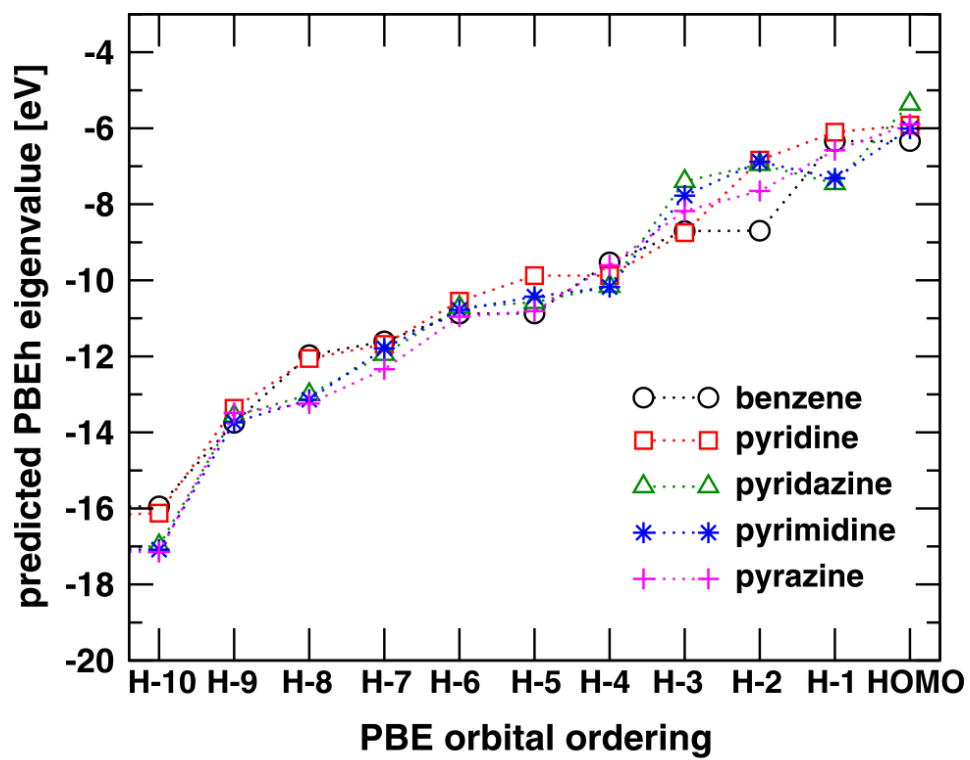
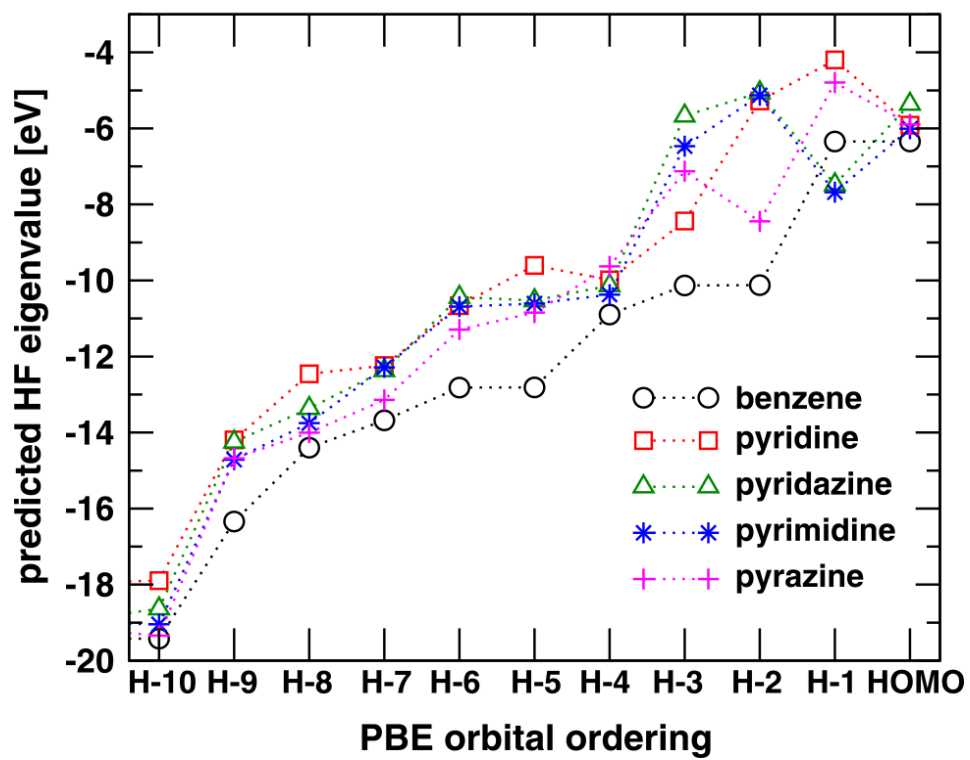

Figure 3: PBEh and HF eigenvalues, as estimated based on a PBE calculation using Eq. (2) and (3), for benzene, pyridine, pyridazine, pyrimidine, and pyrazine.

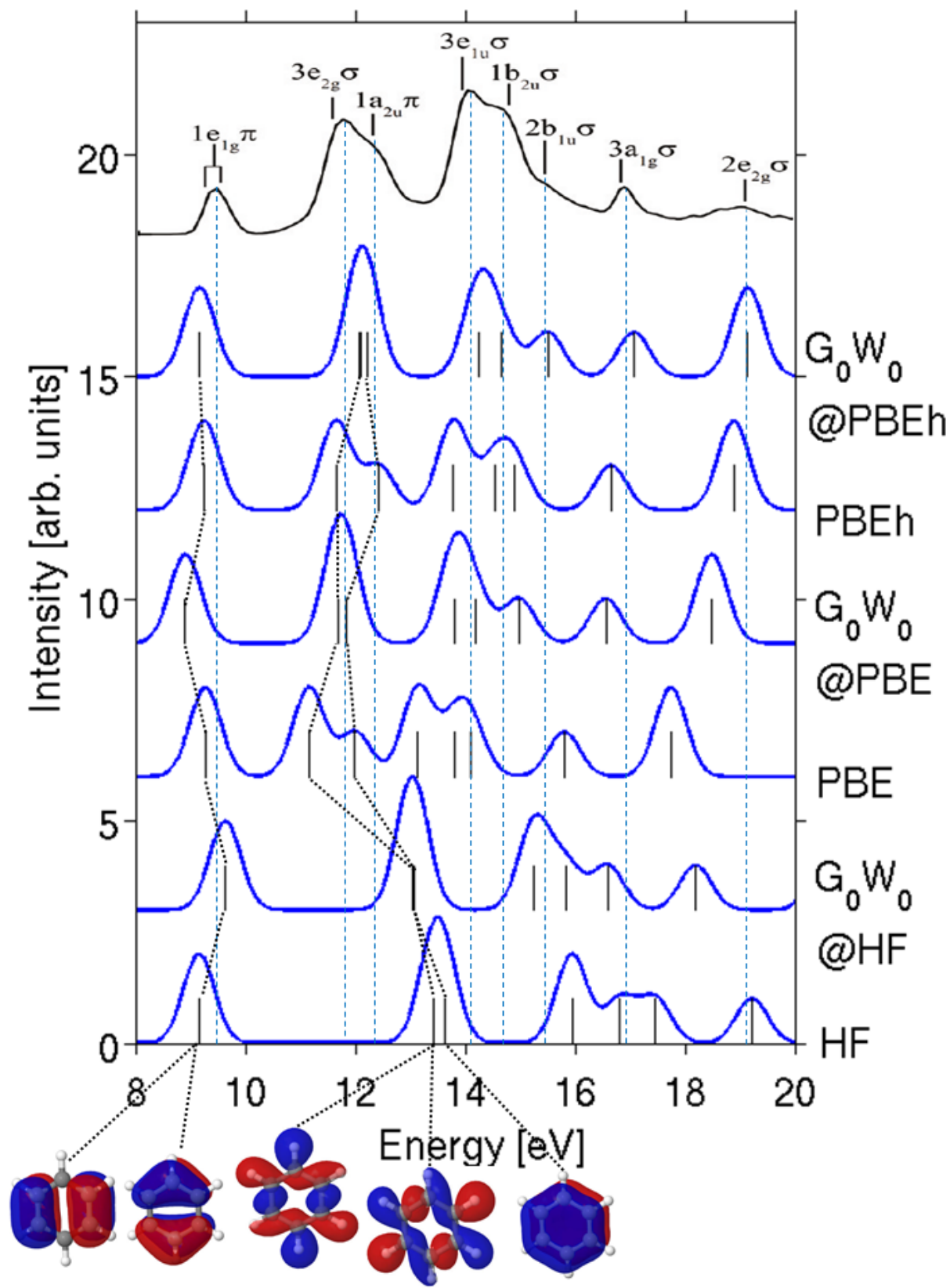

Figure 4: Spectra of benzene, calculated with DFT and $G_0W_0$, broadened by a 0.4 eV Gaussian, compared to gas phase PES.[94] Illustrations of the frontier orbitals are also shown.

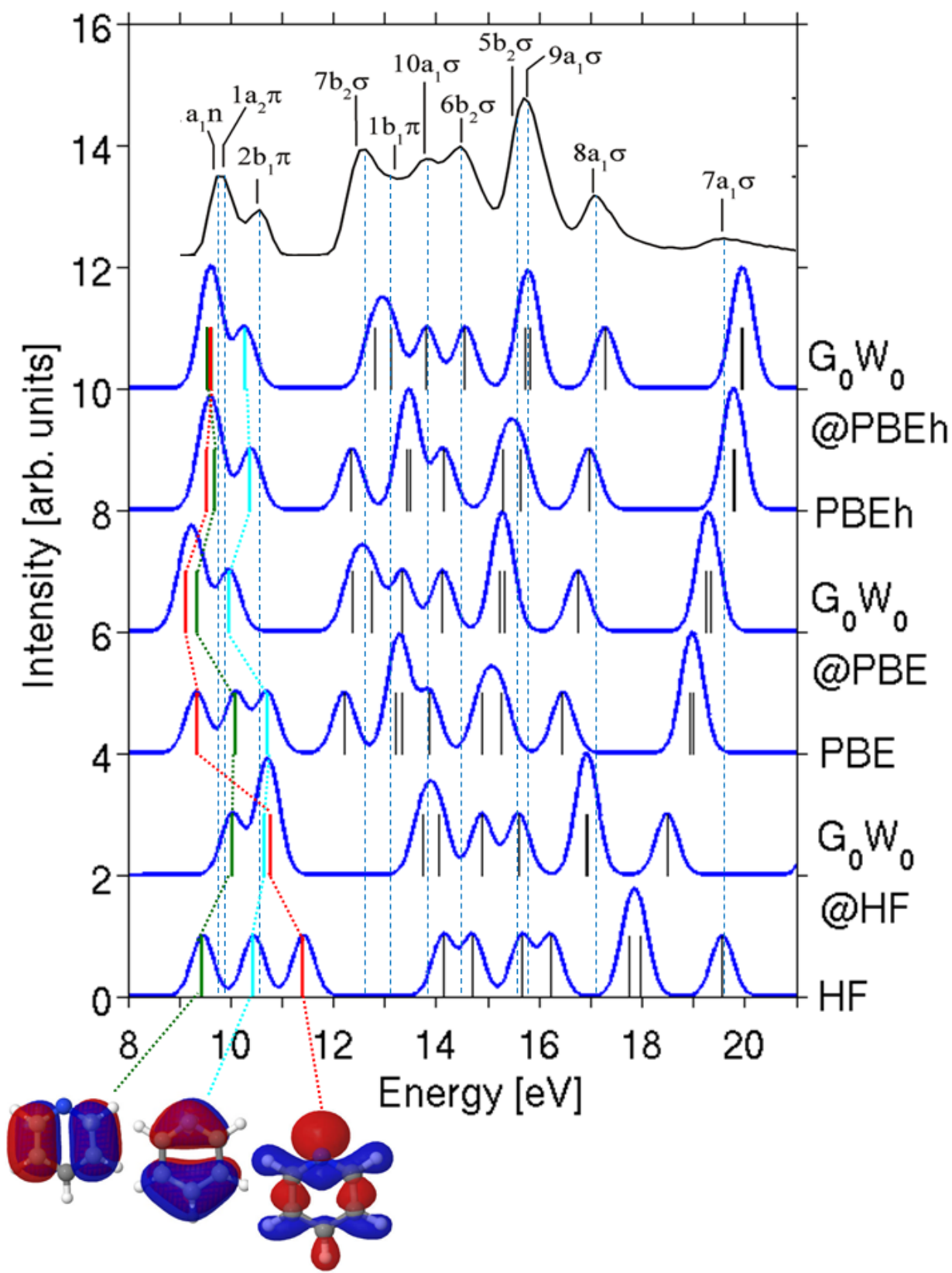

Figure 5: Spectra of pyridine, calculated with DFT and $G_0W_0$, broadened by a 0.3 eV Gaussian, compared to gas phase PES.[94] Illustrations of the frontier orbitals are also shown.

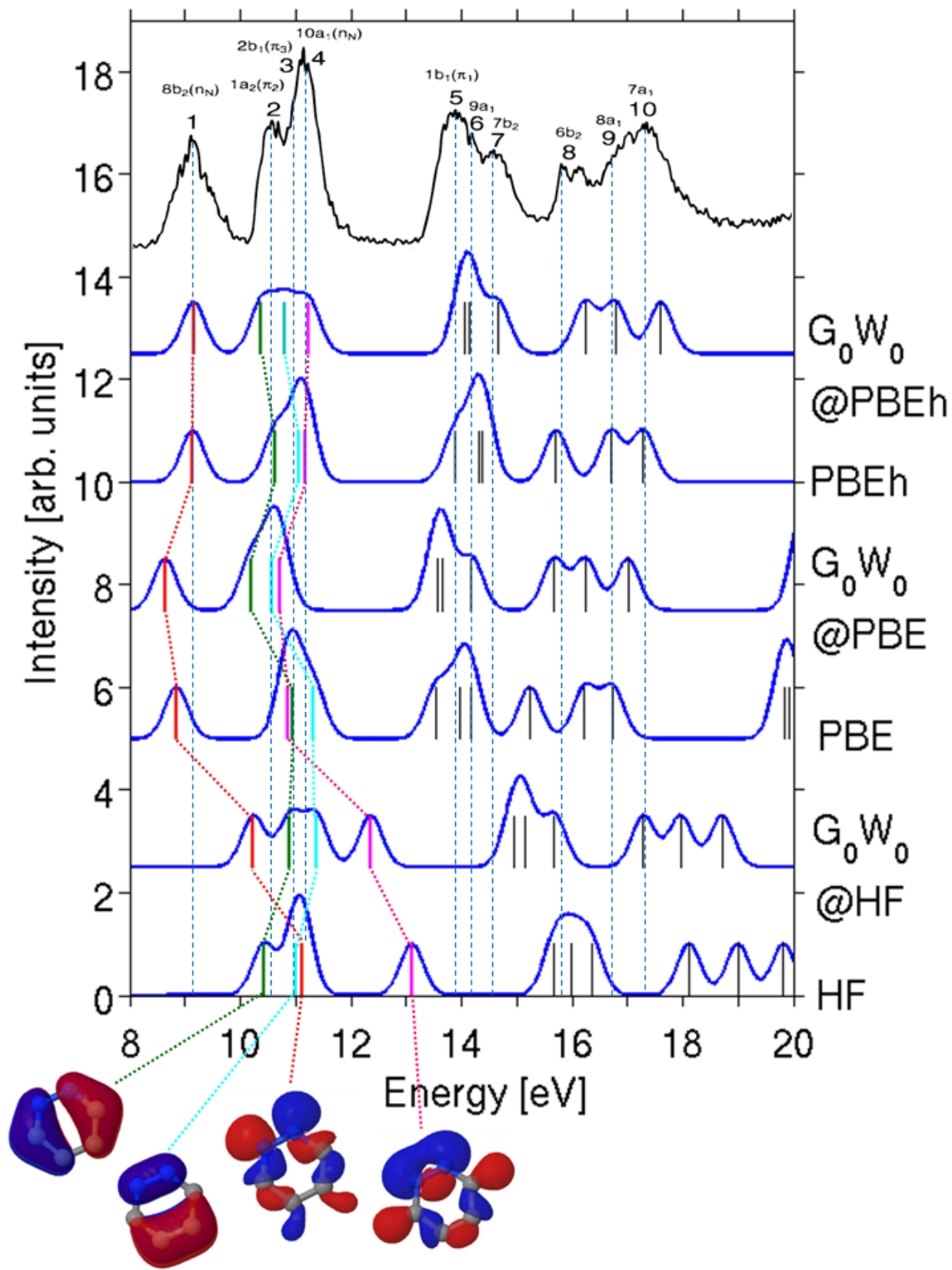

Figure 6: Spectra of pyridazine, calculated with DFT and $G_0W_0$, broadened by a 0.3 eV Gaussian, compared to gas phase PES.[98] Illustrations of the frontier orbitals are also shown.

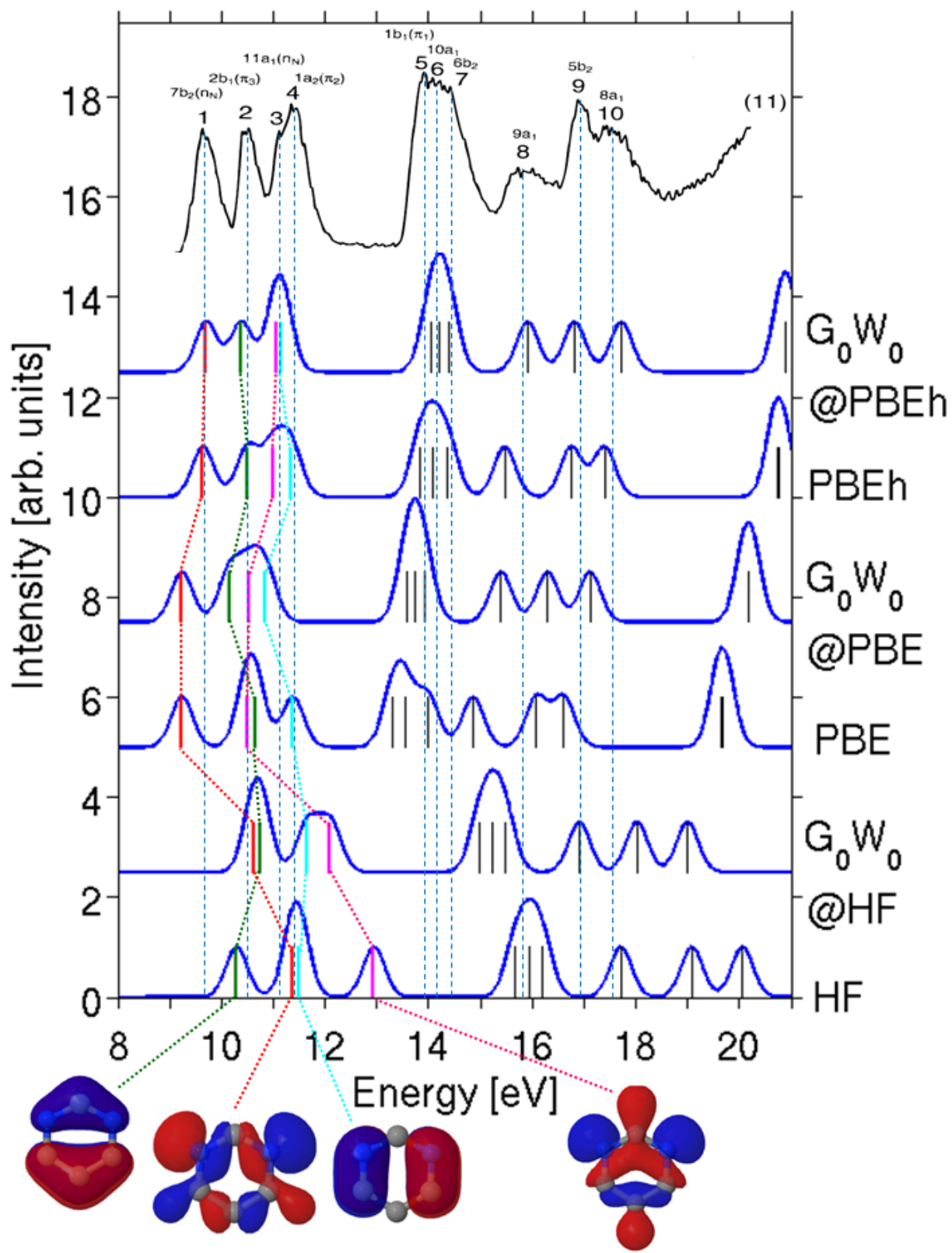

Figure 7: Spectra of pyrimidine, calculated with DFT and $G_0W_0$, broadened by a 0.3 eV Gaussian, compared to gas phase PES.[98] Illustrations of the frontier orbitals are also shown.

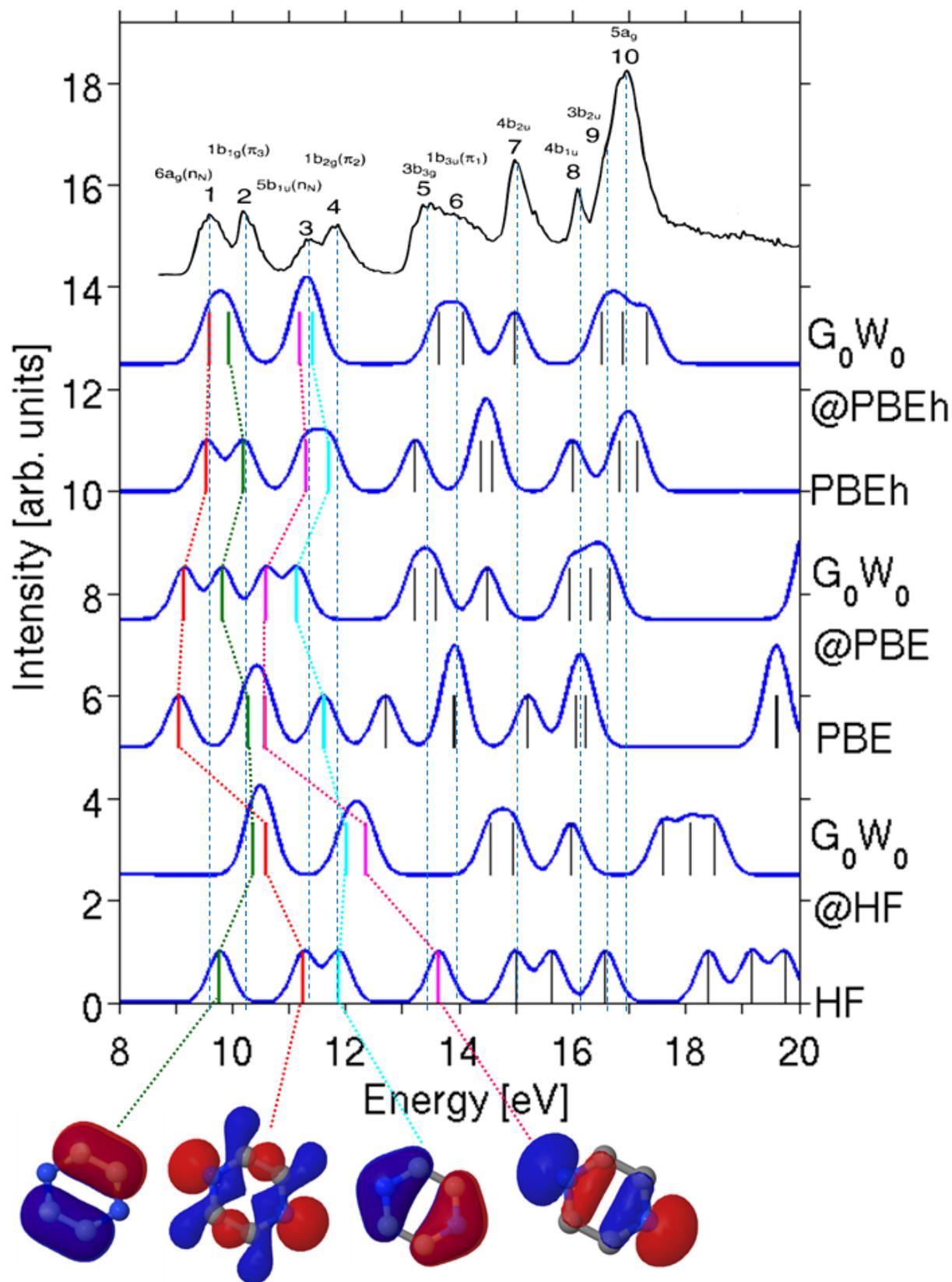

Figure 8: Spectra of pyrazine, calculated with DFT and $G_0W_0$, broadened by a 0.3 eV Gaussian, compared to gas phase PES.[98] Illustrations of the frontier orbitals are also shown.

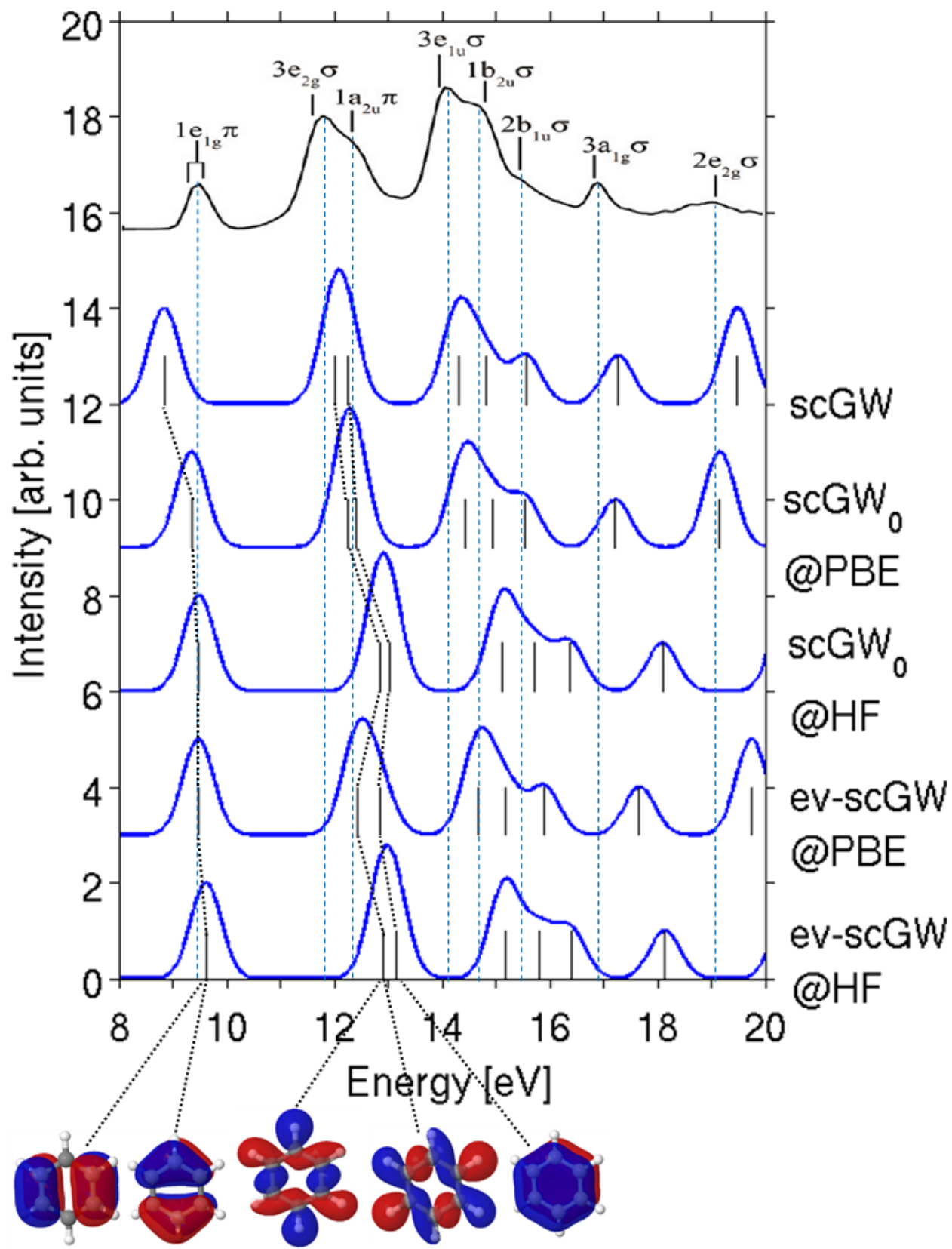

Figure 9: Spectra of benzene, calculated with GW at different levels of self-consistency, broadened by a 0.4 eV Gaussian, compared to gas phase PES.[94] Illustrations of the frontier orbitals are also shown

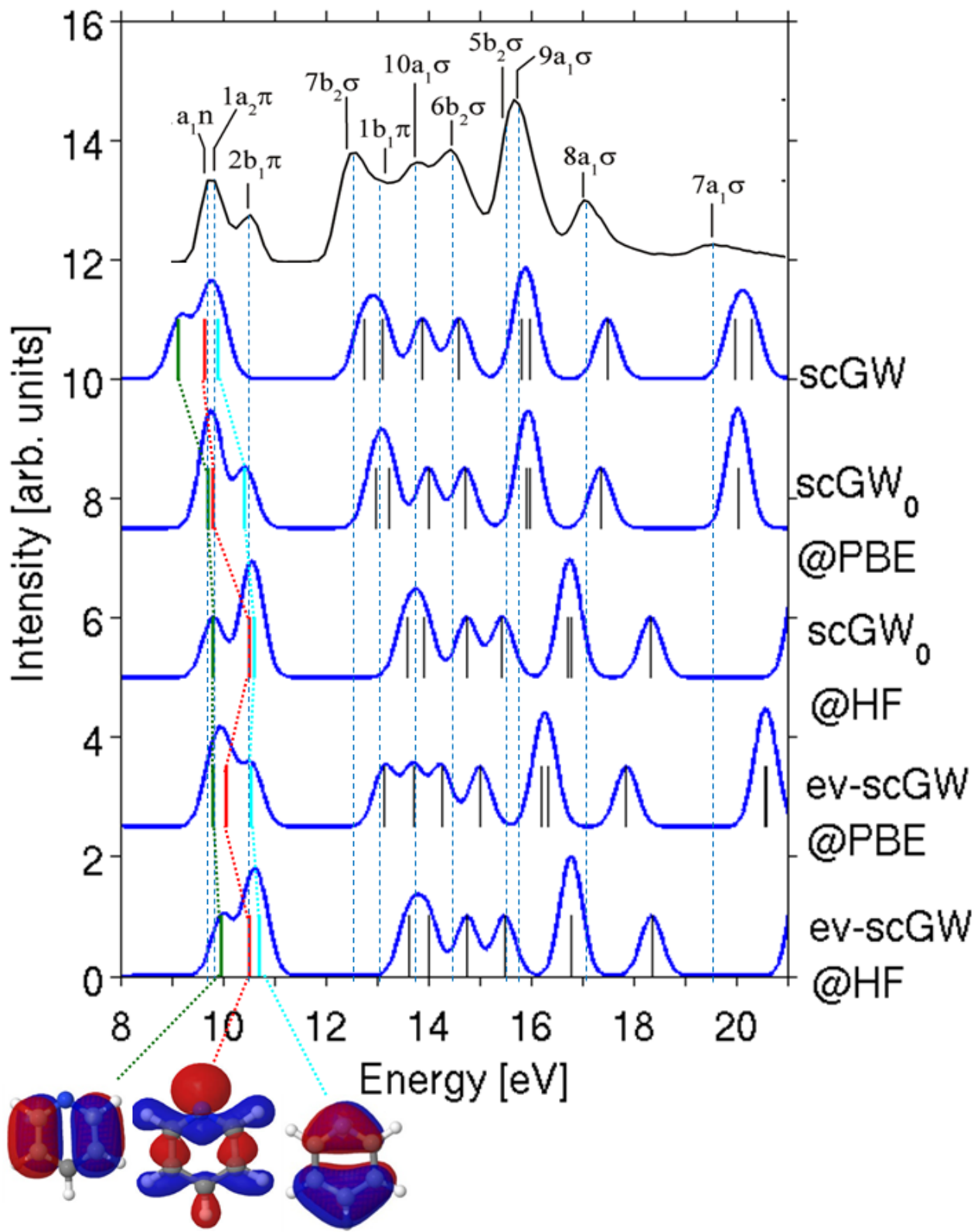

Figure 10: Spectra of pyridine, calculated with GW at different levels of self-consistency, broadened by a 0.4 eV Gaussian, compared to gas phase PES.[94] Illustrations of the frontier orbitals are also shown

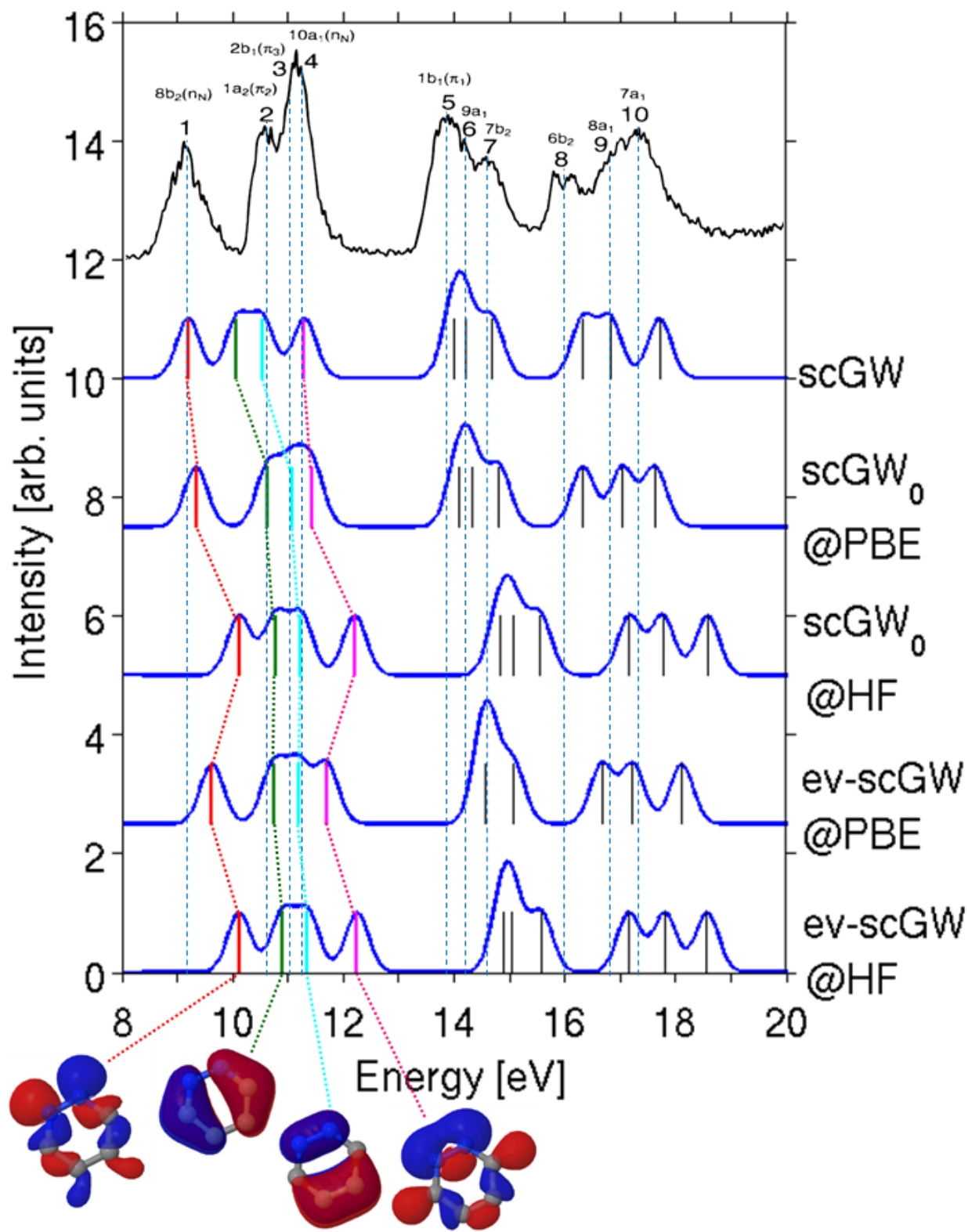

Figure 11: Spectra of pyridazine, calculated with GW at different levels of self-consistency, broadened by a 0.3 eV Gaussian, compared to gas phase PES.[98] Illustrations of the frontier orbitals are also shown.

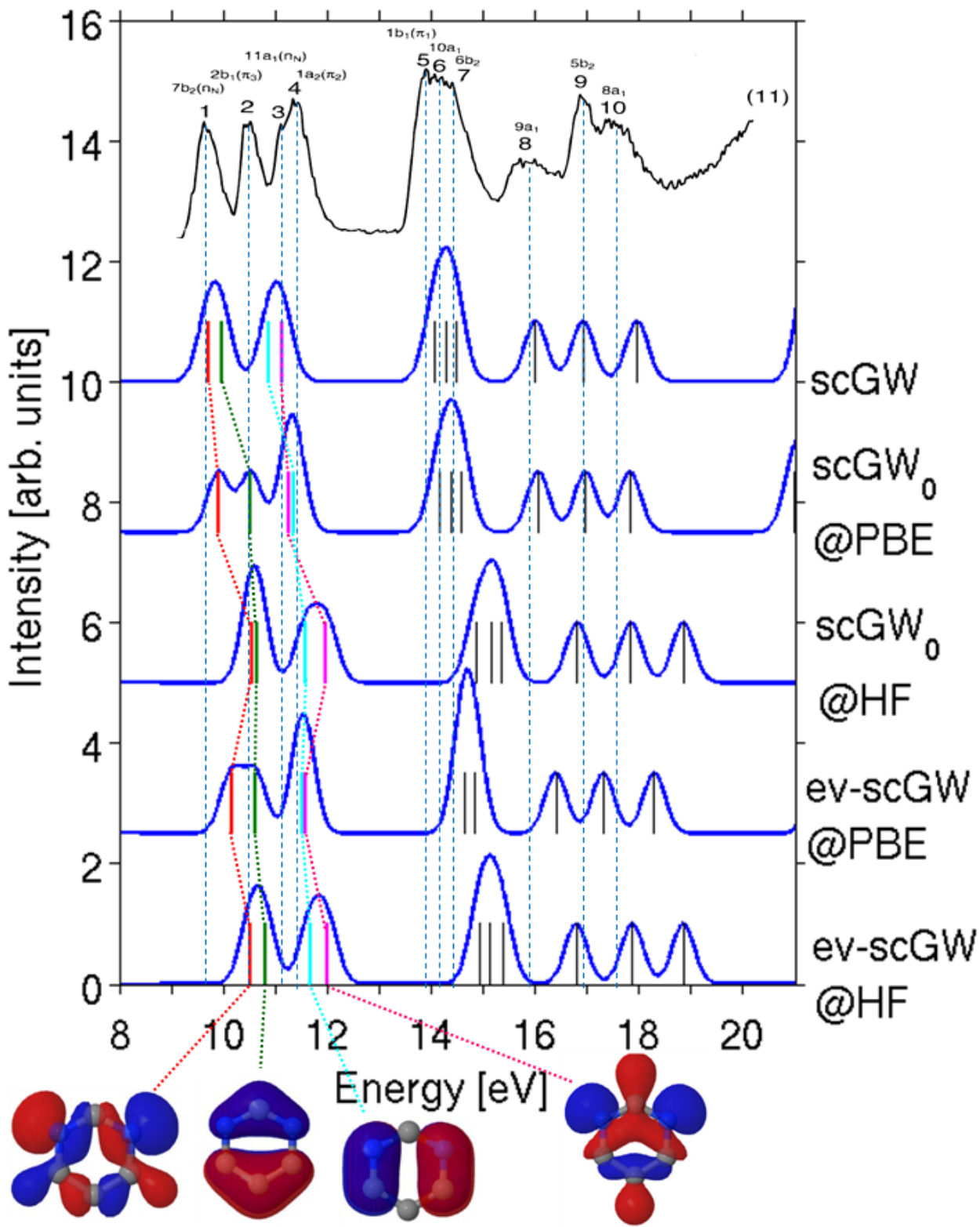

Figure 12: Spectra of pyrimidine, calculated with GW at different levels of self-consistency, broadened by a 0.3 eV Gaussian, compared to gas phase PES.[98] Illustrations of the frontier orbitals are also shown.

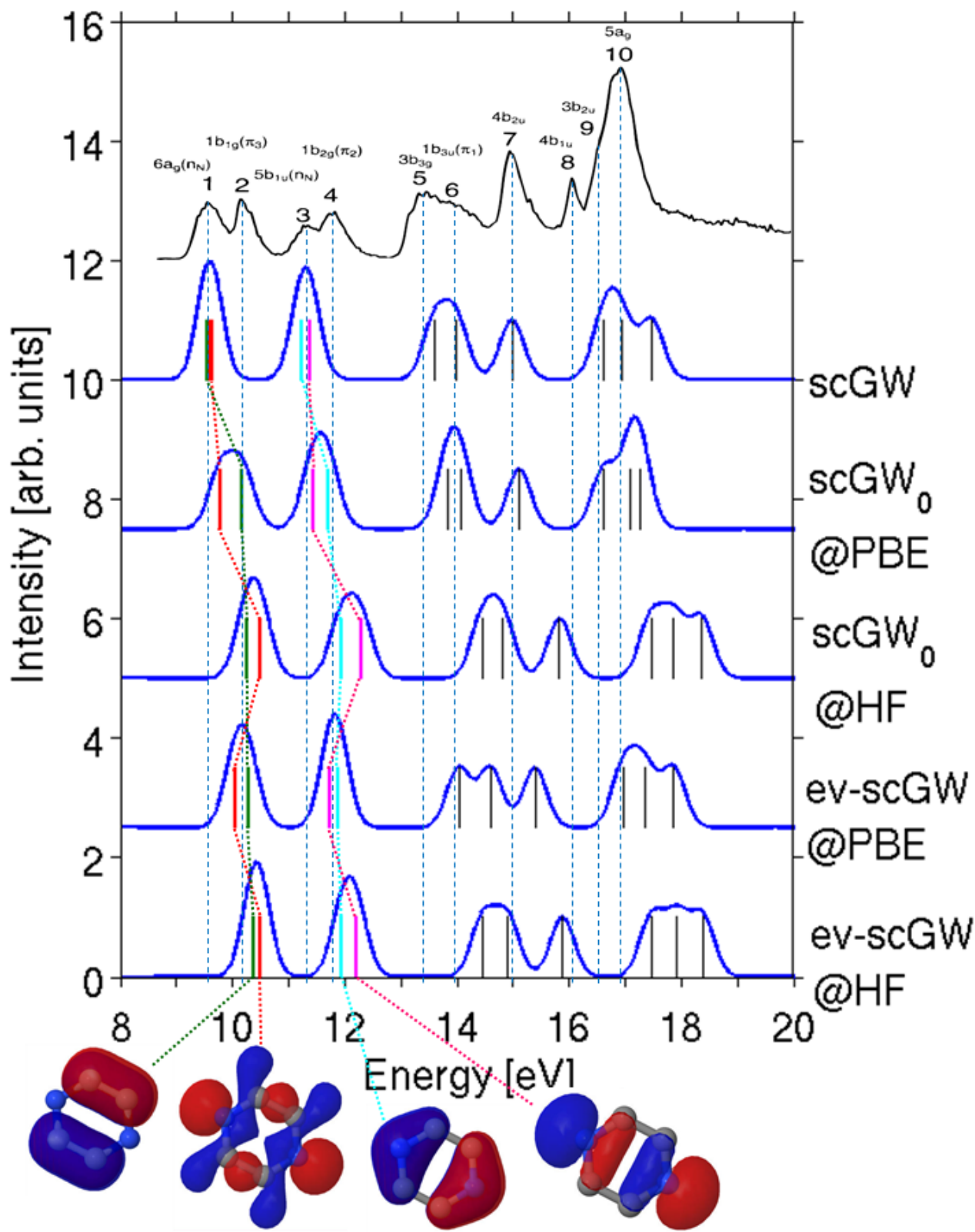

Figure 13: Spectra of pyrazine, calculated with GW at different levels of self-consistency, broadened by a 0.3 eV Gaussian, compared to gas phase PES.[98] Illustrations of the frontier orbitals are also shown.

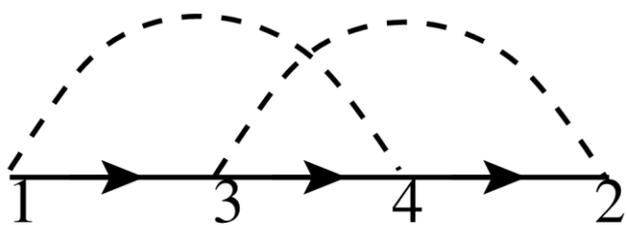

Figure 14: Feynman diagram for the 2nd-order exchange self-energy. Arrows represent the Green's function, and dashed lines represent the (bare) Coulomb interaction.

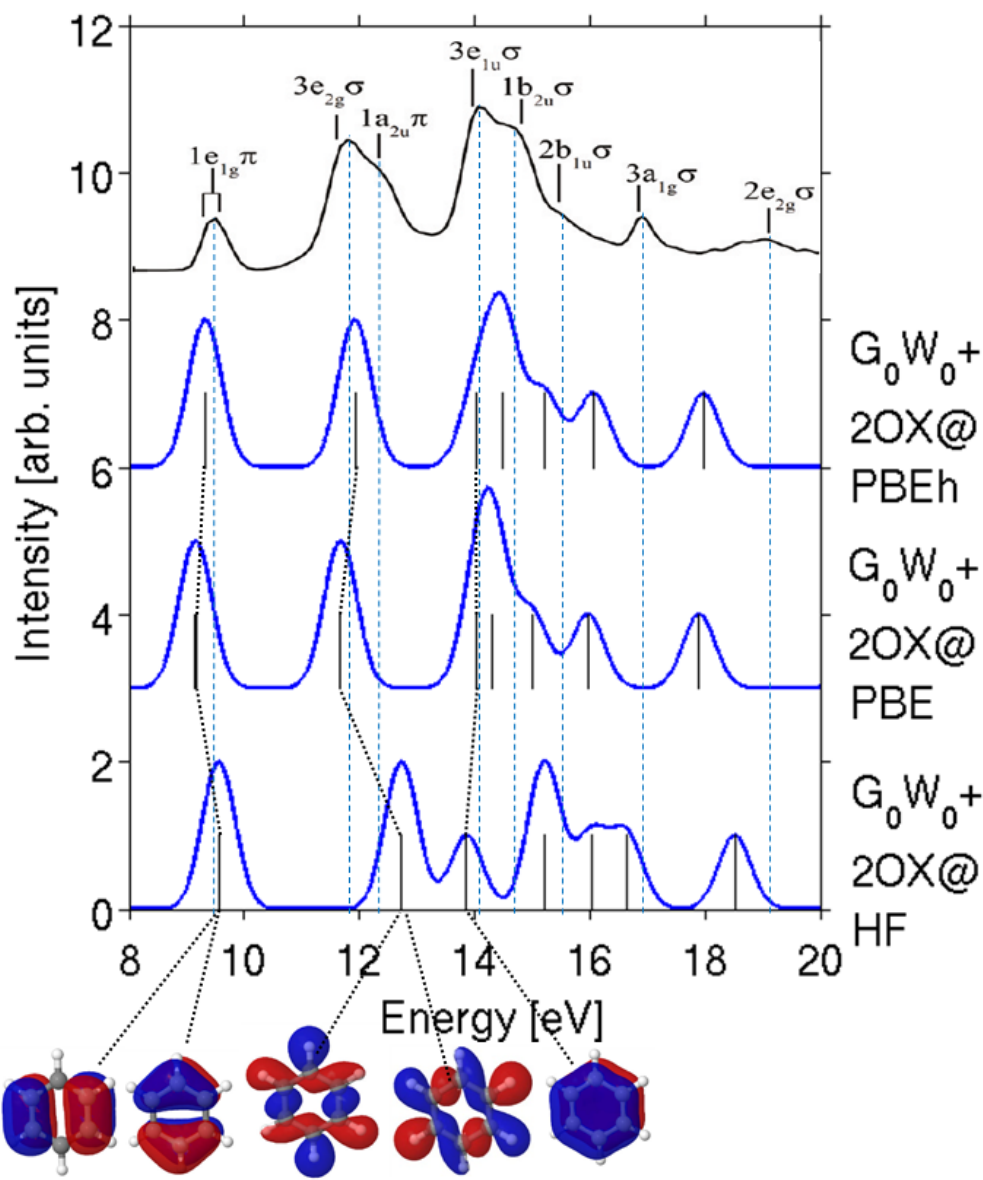

Figure 15: Spectra of benzene, calculated using $G_0W_0$ with second order exchange, based on different DFT starting points, broadened by a 0.4 eV Gaussian, compared to gas phase PES.[94] Illustrations of the frontier orbitals are also shown

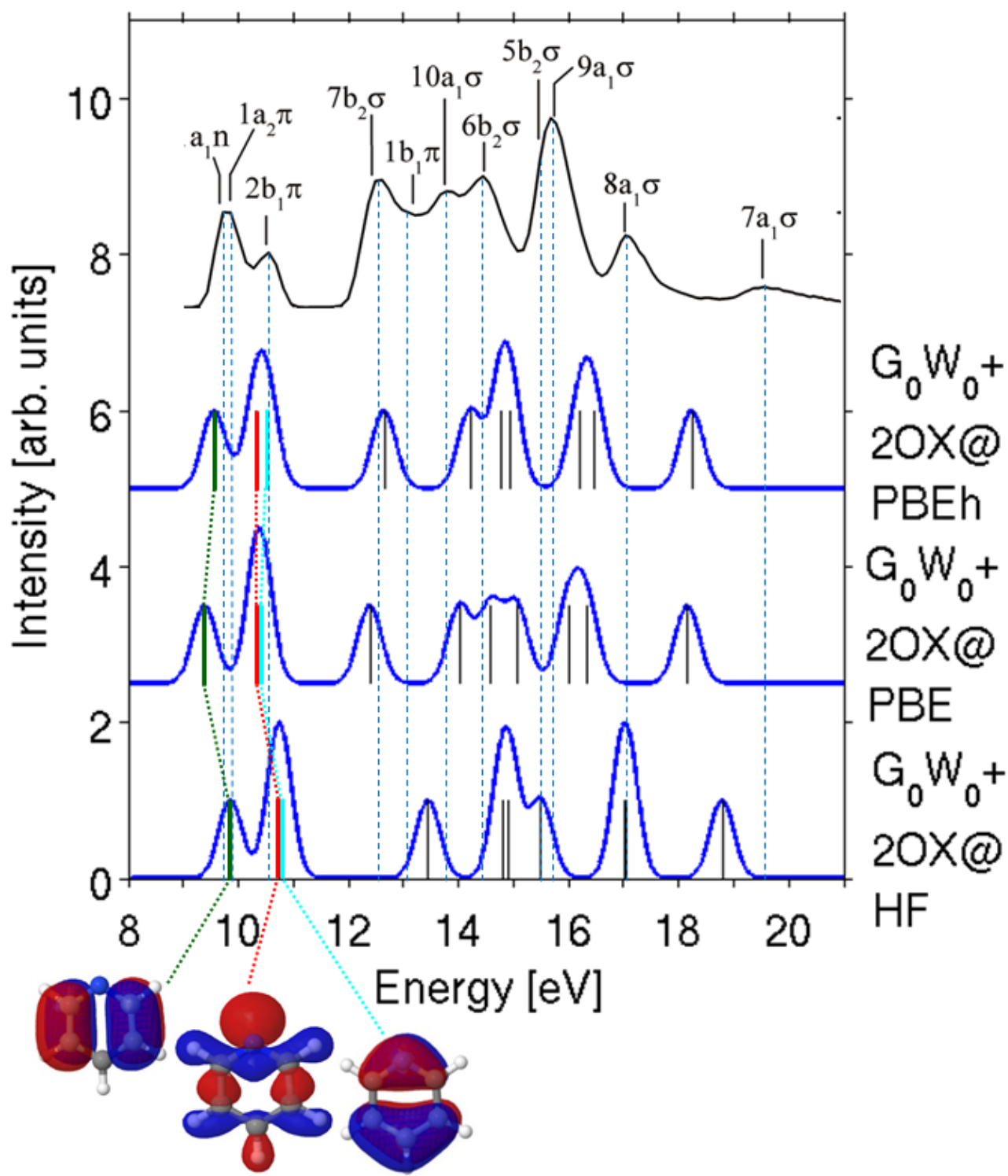

Figure 16: Spectra of pyridine, calculated using $G_0W_0$ with second order exchange, based on different DFT starting points, broadened by a 0.4 eV Gaussian, compared to gas phase PES.[94] Illustrations of the frontier orbitals are also shown

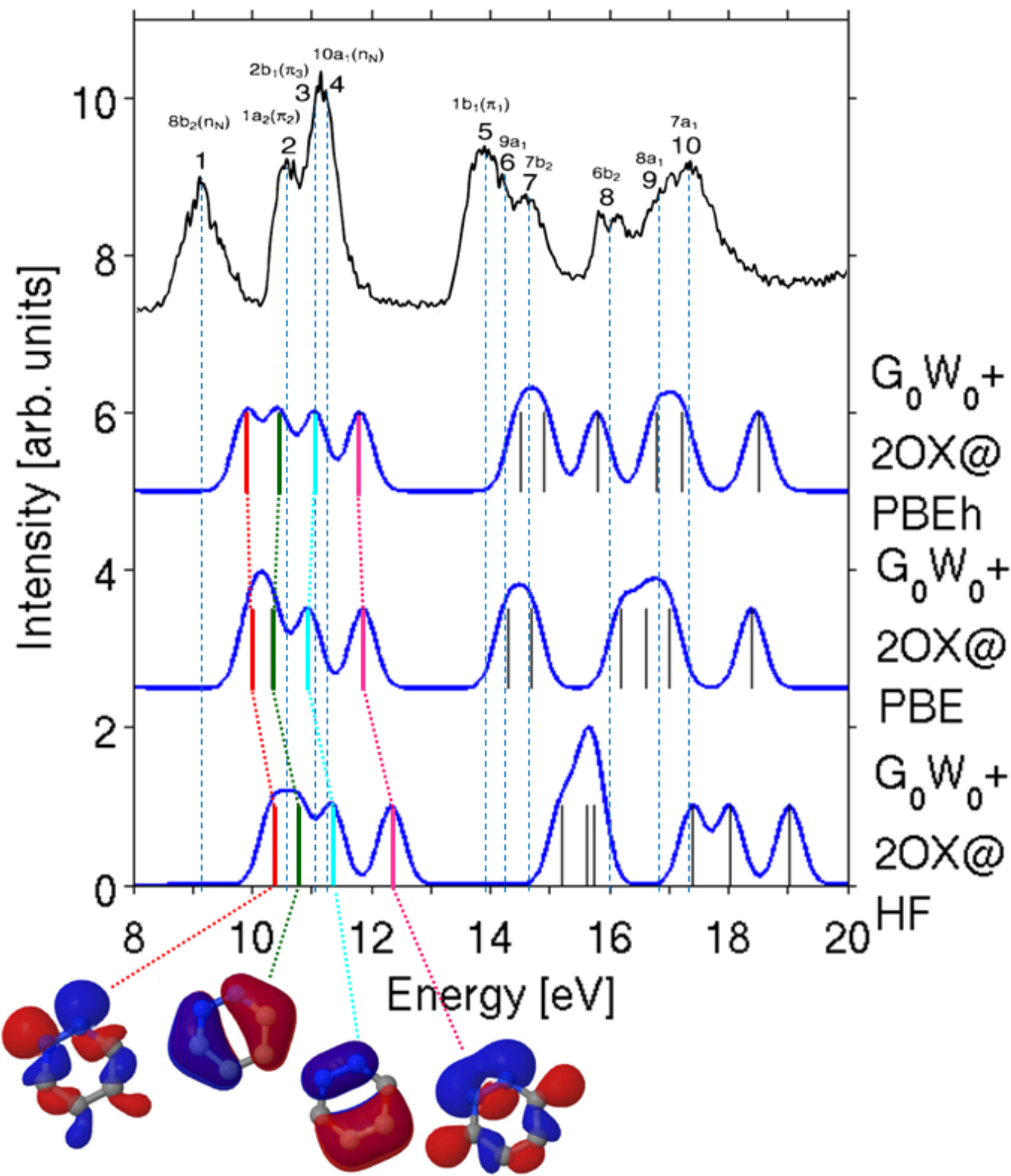

Figure 17: Spectra of pyridazine, calculated using $G_0W_0$ with second order exchange, based on different DFT starting points, broadened by a 0.3 eV Gaussian, compared to gas phase PES.[98] Illustrations of the frontier orbitals are also shown.

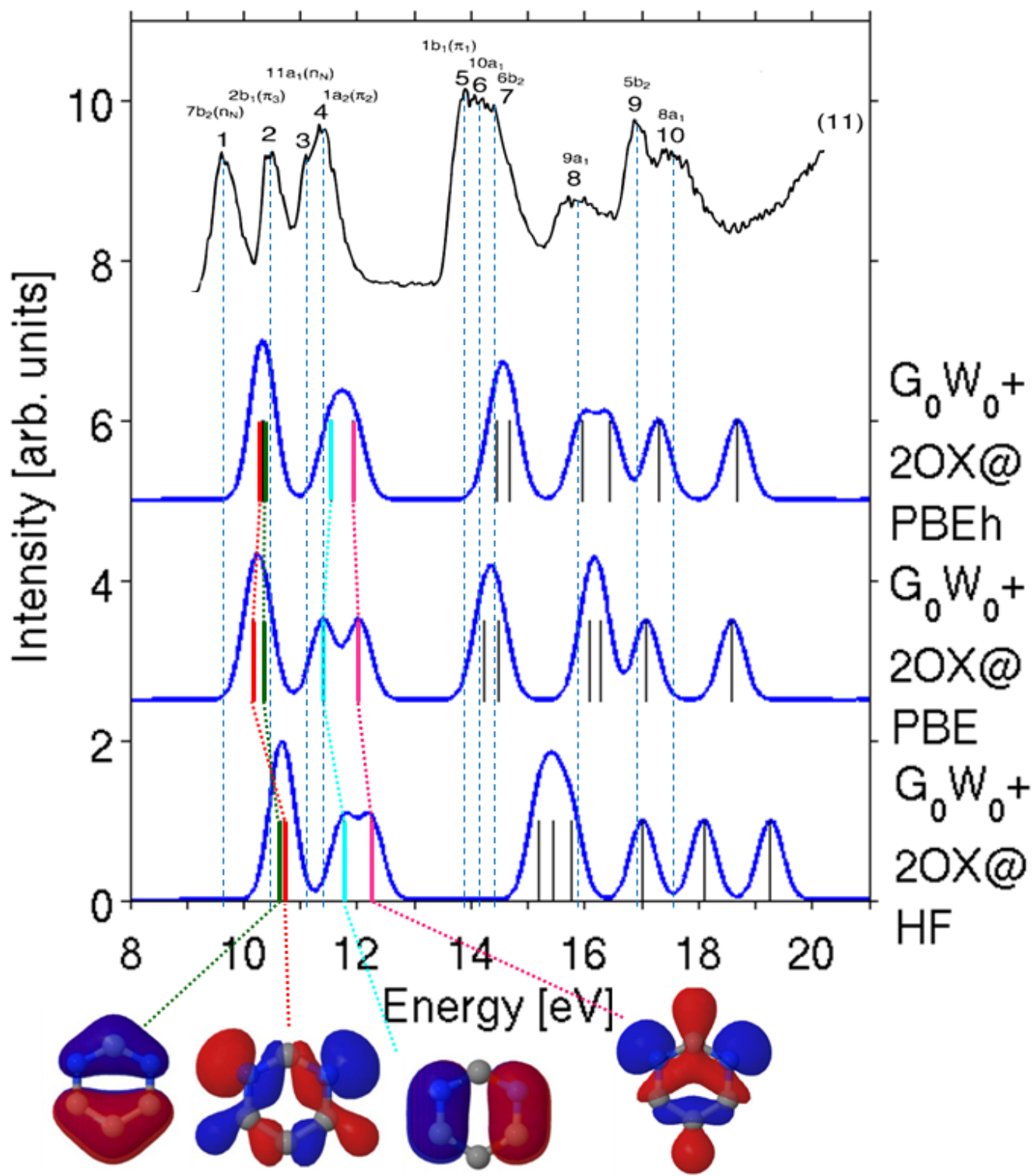

Figure 18: Spectra of pyrimidine, calculated using $G_0W_0$ with second order exchange, based on different DFT starting points, broadened by a 0.3 eV Gaussian, compared to gas phase PES.[98] Illustrations of the frontier orbitals are also shown.

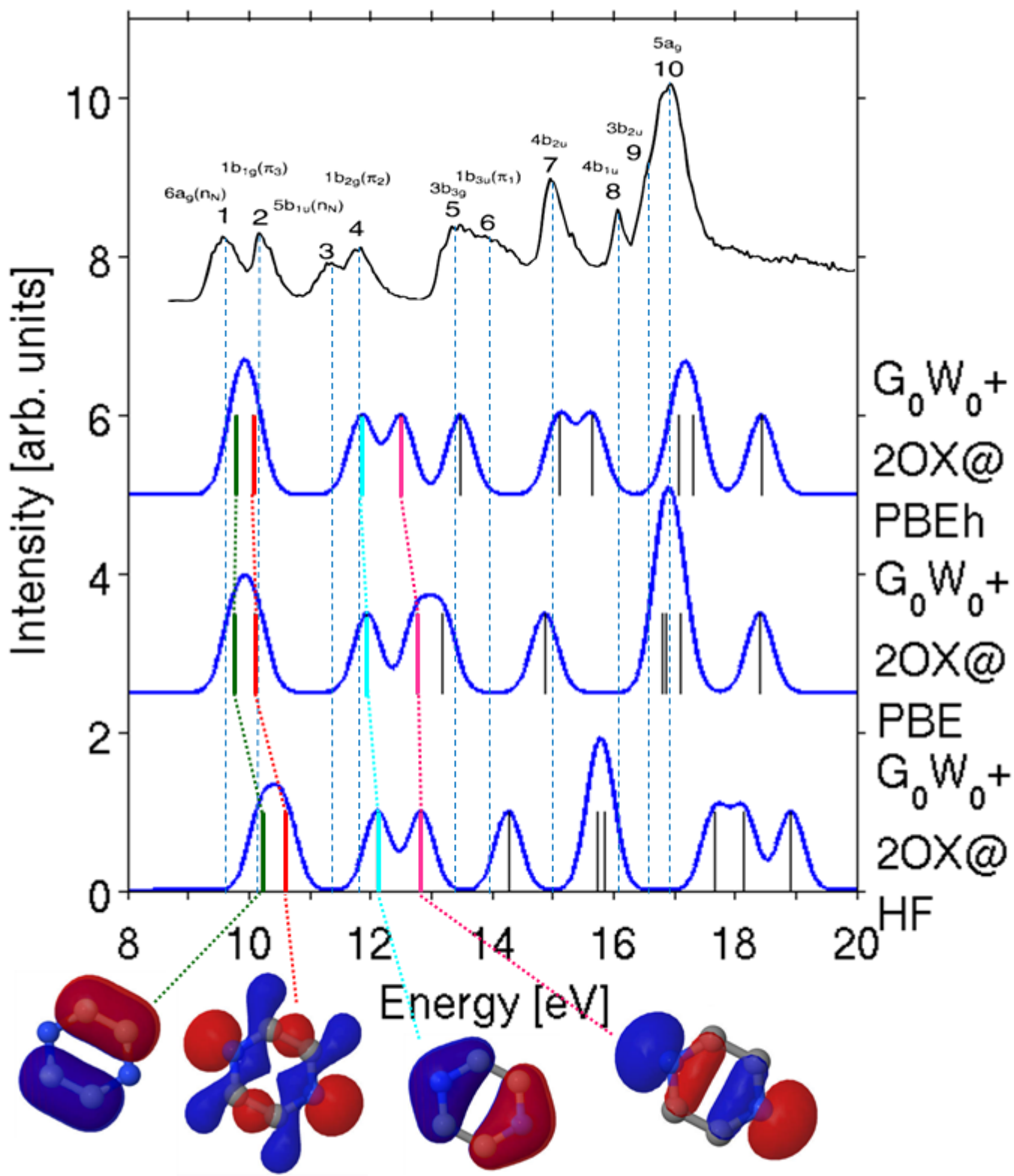

Figure 19: Spectra of pyrazine, calculated using $G_0W_0$ with second order exchange, based on different DFT starting points, broadened by a 0.3 eV Gaussian, compared to gas phase PES.[98] Illustrations of the frontier orbitals are also shown.